\documentclass[iop]{emulateapj}
\usepackage{epsfig,amsmath,natbib}
\usepackage{color}

\def\refjnl#1{{\rm#1}}
\def\aj{\refjnl{AJ}}                   
\def\araa{\refjnl{ARA\&A}}             
\def\apj{\refjnl{ApJ}}                 
\def\apjl{\refjnl{ApJ}}                
\def\apjs{\refjnl{ApJS}}               
\def\aap{\refjnl{A\&A}}                
\def\mnras{\refjnl{MNRAS}}             
\def\nat{\refjnl{Nature}}              
\def\bain{\refjnl{Bull.~Astron.~Inst.~Netherlands}}  

\def\refjnl#1{{\rm#1}}
\def\aj{\refjnl{AJ}}                   
\def\araa{\refjnl{ARA\&A}}             
\def\apj{\refjnl{ApJ}}                 
\def\apjl{\refjnl{ApJ}}                
\def\apjs{\refjnl{ApJS}}               
\def\aap{\refjnl{A\&A}}                
\def\mnras{\refjnl{MNRAS}}             
\def\nat{\refjnl{Nature}}              
\def\bain{\refjnl{Bull.~Astron.~Inst.~Netherlands}}  

\def\be{\begin{equation}} 
\def\ee{\end{equation}} 
\def\ba{\begin{eqnarray}} 
\def\ea{\end{eqnarray}}

\def\cc{\,{\rm {cm^{-3}}}} 
\def\ergs{\,{\rm erg\, {s^{-1}}}}

\def\gsim{\lower.5ex\hbox{\gtsima}} 
\def\lsim{\lower.5ex\hbox{\ltsima}} \def\gtsima{$\; \buildrel > \over 
\sim \;$} \def\ltsima{$\; \buildrel < \over \sim \;$} \def\prosima{$\; 
\buildrel \propto \over \sim \;$} \def\gsim{\lower.5ex\hbox{\gtsima}} 
\def\lsim{\lower.5ex\hbox{\ltsima}} 
\def\simgt{\lower.5ex\hbox{\gtsima}} 
\def\simlt{\lower.5ex\hbox{\ltsima}} 
\def\simpr{\lower.5ex\hbox{\prosima}}   
  
 \def\gtsima{$\; \buildrel > \over \sim \;$} 
\def\ltsima{$\; \buildrel < \over \sim \;$} 
\def\gsim{\lower.5ex\hbox{\gtsima}} 
\def\lsim{\lower.5ex\hbox{\ltsima}} 
\def\simgt{\lower.5ex\hbox{\gtsima}} 
\def\simlt{\lower.5ex\hbox{\ltsima}} 
\def\simpr{\lower.5ex\hbox{\prosima}}

\def\E3{{\cal E}_{\rm g}^{III}}

\def\r12{r_{1/2}} 
\def\x12{x_{1/2}} 
\def\v12{v_{1/2}}

\begin{document}

\title{The Production of Cold Gas Within Galaxy Outflows} 
\author{Evan Scannapieco}
\affil{School of Earth and Space Exploration,  Arizona State University,\\ P.O.  Box 871404, Tempe, AZ, 85287-1404}

  
\label{firstpage} 
\begin{abstract} 

I present a suite of three-dimensional simulations of the evolution of initially-hot material ejected by starburst-driven galaxy outflows. The simulations are conducted in a comoving frame that moves with the material, tracking  atomic/ionic cooling, Compton cooling, and dust cooling and destruction.  Compton cooling is most efficient of these processes, while the main role of atomic/ionic cooling is to enhance density inhomogeneities.  Dust, on the other hand, has little effect on the outflow evolution, and is rapidly destroyed in all the simulations except the case with the smallest mass flux. I use the results to construct a simple steady-state model of the observed UV/optical emission from each outflow.  The velocity profiles in this case are dominated by geometric effects, and the overall luminosities are extremely strong functions of the properties of the host system, as observed in ultra-luminous infrared galaxies (ULIRGs).  Furthermore the luminosities and maximum velocities in several models are consistent with emission-line observations of ULIRGs, although the velocities are significantly greater than observed in absorption-line studies.  It may be that absorption line observations of galaxy outflows probe entrained cold material at small radii, while emission-line observations probe cold material condensing from the initially hot medium at larger distances.

\end{abstract}


\section{Introduction}
\label{Int}

Many of the observed features of galaxies are thought to be caused by outflows. The material they transport helps to determine the number density and baryonic content of dwarf galaxies \citep[e.g.][]{2000MNRAS.319..168C,2001ApJ...557..605S,SFM02,Bens+03}; the metals they remove helps to determine the relationship between the interstellar medium metallicity and galaxy stellar mass,  \citep[e.g.][]{2004ApJ...613..898T, 2008ApJ...681.1183K,2010MNRAS.408.2115M}; and the enriched material they disperse helps to determine the metallicity history of the intergalactic medium  \citep[e.g.][]{2005ApJ...634L..37F, 2010ApJ...721..174M, 2014ApJ...786...54P, 2015MNRAS.450.2067T}.  Yet despite their importance, the dynamical and microphysical processes that control galaxy outflows remain extremely poorly understood.

From a theoretical perspective, the most important outstanding issue is that the highly efficient cooling within the interstellar medium (ISM) makes it impossible to model supernovae by simply adding thermal energy to the medium, unless both the Sedov and shell formation stages are sufficiently well resolved \citep{2015ApJ...802...99K}.  At the same time, currently-available computational resources  do not allow for direct modeling of supernovae within a galaxy-scale simulation \citep[e.g.][]{2004A&A...425..899D,2011ApJ...731...11C, 2013MNRAS.430.3213B, 2015MNRAS.446.2125C, 2015MNRAS.454..238W}.  The connection to observations is also complicated by the possibility that radiation pressure on dust \citep{2005ApJ...630..167T, 2011ApJ...735...66M, 2011MNRAS.417..950H}, non-thermal pressure from cosmic rays \citep{2008ApJ...687..202S,2012MNRAS.423.2374U,2013ApJ...777L..16B,2014MNRAS.437.3312S}, and energy input from gravitationally driven motions \citep{2016ApJ...818...28S} may also contribute to driving winds.

From an observational perspective, the most important issue is that the multiphase nature of galaxy outflows causes us to only have a partial view of the material they eject.  In fact the outflowing material spans an enormous range of  temperatures: from $10^7-10^8$K plasma observed in X-rays  \citep[e.g.][]{1999ApJ...513..156M,2007ApJ...658..258S,2009ApJ...697.2030S}, to $\approx10^4 $K ions observed at optical and near UV wavelengths   \citep[e.g.][]{2001ApJ...554..981P,2007ApJ...663L..77T,2012ApJ...760..127M,2012ApJ...757...86S}, to $10-10^3$ K molecular gas observed at radio wavelengths \citep[e.g.][]{2002ApJ...580L..21W,2011ApJ...733L..16S,2013Natur.499..450B}.   Furthermore, the easiest phase to interpret, the  X-ray emitting gas, is detectable only in the nearby Universe. The colder phases, on the other hand, can be easily observed out to high redshifts,  and their kinematics have been carefully studied in many cases \citep[e.g.][]{2004ApJ...610..201S, 2005ApJS..160..115R, 2012ApJ...759...26E, 2009ApJ...692..187W, 2013ApJ...774...50K,2015ApJ...809..163A}.  On the other hand, not only is the connection between the motion of this material and the X-ray emitting gas highly uncertain, but its overall origin remains unknown. 

Supernovae are not evenly distributed in time and space, and so complete thermalization of the outflowing medium may not be achieved, in particular when the star formation rate is small and supernova events not so frequent.  Incomplete thermalization would cause the outflow to be born as a collection of hot gas surrounding cold clouds that would then be driven out of the host galaxy by ram pressure acceleration from hotter material \citep[e.g.][]{Veilleux2005}.   However, this model runs into serious difficulties in explaining the observations, both because: (i) shocks and conduction from the exterior medium tend to compress the cloud perpendicular to the direction of the flow, greatly reducing the momentum flux it receives; and (ii) instabilities and evaporation lead to rapid cloud disruption  \citep{Klein94,1994ApJ...434L..33M,2006A&A...457..545O,2008ApJ...678..274O,Scannapieco15,2016ApJ...822...31B}.   Together these imply that the lifetimes of the clumps are likely to be much shorter that the time required to accelerate them to the observed speeds.

A second possibility is that thermalization of the outflowing medium is efficient and the cold clouds are instead formed later, directly from the cooling of high temperature ($10^7-10^8$K) material, already moving at high radial velocities.  This was first proposed by  \cite{1995ApJ...444..590W,1995ApJ...444L..17W}, who calculated steady-state one-dimensional profiles of outflowing winds including cooling, parameterized as a power law.   A more detailed one-dimensional model was then developed by \cite{2000MNRAS.317..697E}, which incorporated the infall of cooling gas from a halo, and the outflow of hot gas from the interstellar medium.  \cite{2003ApJ...590..791S, 2004ApJ...610..226S}, \cite{2007ApJ...658.1196T}, and \cite{2011ApJ...740...75W} presented steady-state one and two-dimensional hydrodynamic calculations of galaxy outflows including radiative cooling, accounting for chemical evolution, and comparing their results with observations of several nearby galaxies.  \cite{2015ApJ...803....6M} studied simple models of cooling winds in the context of the Ly$\alpha$ line profiles of ultra-luminous infrared galaxies (ULIRGs).  Most recently, \cite{2016MNRAS.455.1830T} derived one-dimensional models of wind evolution over a large parameter space, including a radiative cooling function that captured the key role of metal-line cooling at temperatures below $10^7$ K \citep[see also][]{Zhang2015}. They also provided scalings for the radius at which gas cools efficiently as a function  density, column density, emission measure, radiative efficiency, and cool gas velocity, highlighting the  dependence of the results on the mass entrained by the wind and  the host galaxy star formation rate surface density.

Here I present three-dimensional simulations of the evolution of cooling gas in galaxy outflows that include several physical processes beyond those considered in previous studies.  To maximize the resolution of the calculations, I adopt a comoving frame that moves radially outward with the outflowing material, capturing both the global cooling of the flow, and the formation of clumps within it.  The cooling routines make use of the code capabilities developed in \cite{2016arXiv160902561F}  to study quasar-driven outflows, including: (i) radiative line cooling that fully accounts for the impact of the ionizing background  and the additional heating due to phoionization, (ii) cooling due to the Compton scattering between free elections and photons from the host galaxy, which can exceed metal-line cooling by over an order of magnitude at small radii, and (iii)  cooling from dust grains, calculated by self-consistently tracking the destruction of dust by sputtering by the hot outflow material.

The structure of this paper is as follows.  In \S2, I describe the simulation methods, focusing on the approach used to work in a comoving frame,  the modeling of atomic and ionic cooling and Compton cooling, and the modeling of dust cooling and destruction.  In \S3, I present the simulation results, describing first the impact of the various physical processes on a fiducial case, and then contrasting this evolution with outflows with varying energy input rates, mass loading factors, and star formation rates.   In this section, I also examine the size distribution of the cold clumps produced in the simulations, and in \S4, I use these results to assess the observability of the cold gas in each of the cases studied. Conclusions are given in \S5.
 
\section{Methods}

To study the formation of cold clouds in starburst-driven outflows,  I carried out a set of simulations using FLASH (version 4.2), a multidimensional hydrodynamic code \citep{2000ApJS..131..273F}.  The simulations used an unsplit hydrodynamic solver with a predictor-corrector type formulation \citep{2013JCoPh.243..269L}, and they made use of the shock-detect flag which lowered the prefactor in the Courant-Friedrichs-Lewy timestep condition from 0.4 to 0.25 in the presence of strong shocks.  The simulations included atomic / ionic cooling and heating process, Compton cooling, and dust cooling and destruction using the methods described in detail in \cite{2016arXiv160902561F}, modified to handle the current problem as described below.    Unlike the simulations in \cite{2016arXiv160902561F} however,  I also modified the FLASH code follow a cubic periodic volume in a comoving frame that allows the medium to expand, conserving mass, as it moves outward from the galaxy.  Here I describe each of the physical processes included in the calculations, beginning with this new approach.

\begin{figure*}[t]
\center{\includegraphics[width=5.5in]{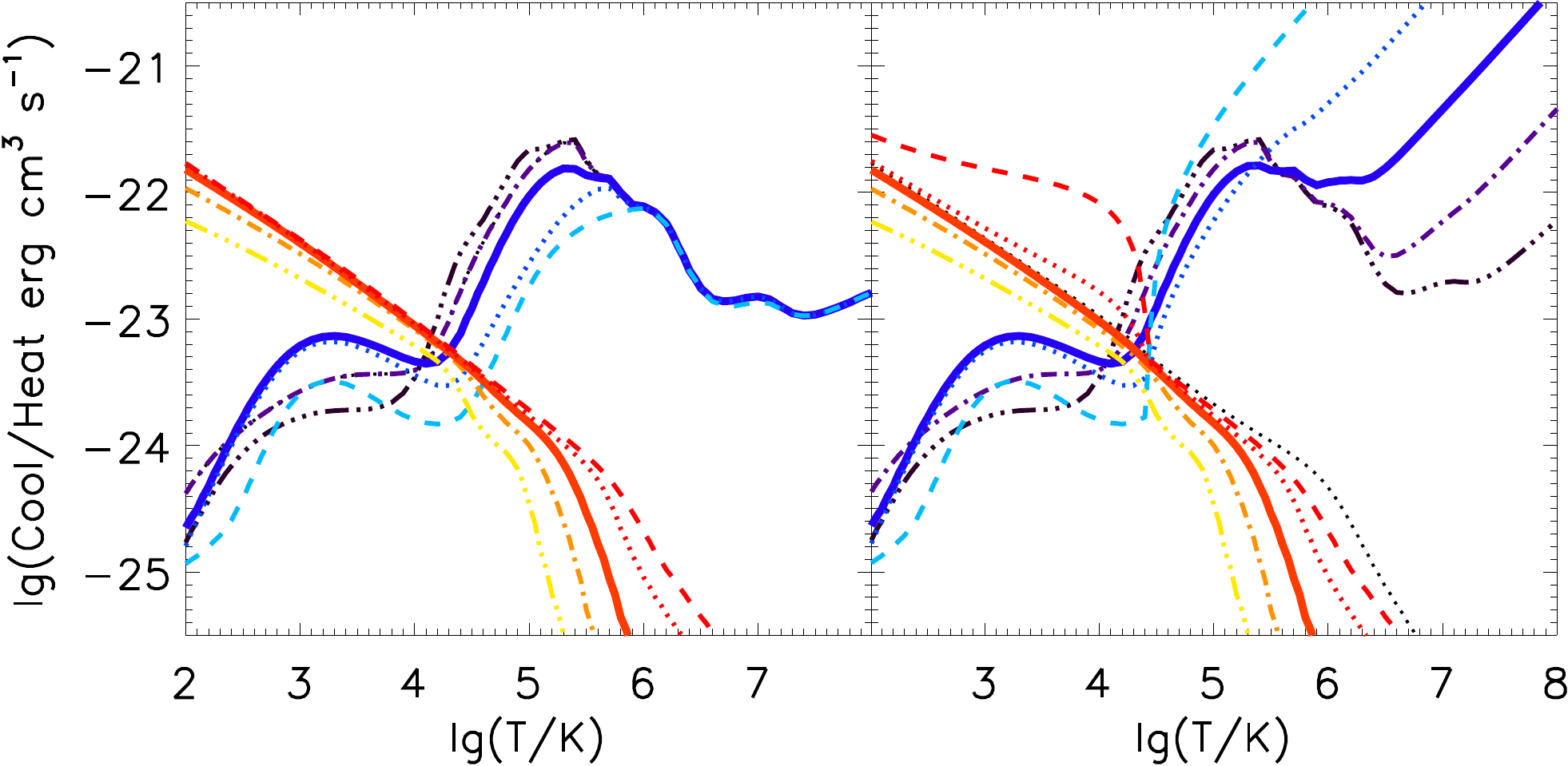}}
\caption{ {\em Left:} Radiative cooling rates (blue lines)  vs. photoheating rate (red lines) for optically thin material, normalized by the mean baryonic density squared.     {\em Right:}  Radiative cooling rates and heating ratios including Compton heating/cooling as given by eq.\ (\ref{eq34a}). In all panels, the solid lines are for material with a mean density of $\rho/\mu m_p$ = 1 cm$^{-3}$, while the dashed, dotted, dot-dashed, and triple-dot-dashed lines correspond to material with $\rho/\mu m_p$ = 0.01, 0.1, 10, and 100 cm$^{-3}$, respectively, for a medium at a distance of 300 parsecs from a starburst with a SFR of 10 $M_\odot$ yr$^{-1}$ \vspace{0.18in}} 
\label{Fig:cool}
\end{figure*}

\subsection{Comoving Frame}

We are interested in gas condensing within a thermalized hot wind expanding rapidly from a starbursting galaxy. For most galaxies, this hot medium moves many scale heights per Myr, while a typical starburst episode  lasts for many Myrs \citep[e.g.][]{1998ApJ...504..725G,2003ApJ...599..193F}. Thus, the distribution is expected to be well-approximated by an equilibrium configuration, and this seems to be born out observationally in most objects in which reliable X-ray analyses can be made \citep[e.g.][]{1990ApJS...74..833H,1995ApJ...448...98H,2005MNRAS.358.1453O,2007ApJ...658..258S,2012ApJ...758..105Y}. In this case, assuming a conical expansion with a fixed opening angle, the equations of mass, momentum, and energy conservation are:
\be
\frac{1}{r^2} \frac{d}{dr} (\rho v r^2) = \dot q_{\rm m},
\label{eq:mass}
\ee
\be
\rho v \frac{dv}{dr} = - \frac{dP}{dr} - \dot q_{\rm m} v,
\label{eq:momentum}
 \ee
\be
\frac{1}{r^2} \frac{d}{dr} \left[ \rho v r^2 \left( \frac{v^2}{2} + \frac{\gamma}{\gamma -1} \frac{P}{\rho} \right) \right] =  \dot q_{\rm e},
\label{eq:energy}
\ee
where $\rho$, $v$, $P,$ and $\gamma$ are the density, radial velocity,  pressure, and the ratio of specific heats of the outflow and the mass and energy input rates are
\ba
\dot q_{\rm m} &=  &
\begin{cases}
\dot q_{\rm m,0} &  \qquad  \qquad  \, \,\, \, \text{if } r \leq r_\star \\
   0       &  \qquad \qquad  \, \,\, \, \text{if } r > r_\star,
\end{cases} \nonumber  \\
\dot q_{\rm e} &=  &
\begin{cases}
  \dot q_{\rm m,0}  c_{s,0}^2 / (\gamma-1) & \text{if } r \leq r_\star \\
  - \Lambda n^2       & \text{if } r >r_\star,
\end{cases}
\ea
respectively, where $r_\star$ is the driving radius of the flow, $c_{s,0}$ is the sound speed of the hot medium at $r =0,$ 
$\Lambda$ is the radiative cooling function, and $n = \rho/\mu m_p$ is the total number density of the medium, with $\mu$ the average particle mass in units of the proton mass, $m_p$.
For reference, in the case of M82, $r_\star \approx 300$ pc \citep{1995A&A...293..703M,2009ApJ...697.2030S}.
Note that here I have ignored radiative cooling within the driving radius, including this in the definition of $c_{\rm s,0}$.

The solution to these equations is given by the \cite{1985Natur.317...44C} model, which can be used to gain a good understanding of the hot wind conditions as a function of radius.  For a $\gamma = 5/3$ gas, the sound speed  when $r < r_\star$ is approximately constant at 
\be
c_{s,{\rm hot}}(r) \approx c_{s,{\rm hot},0} = 0.82 \left( \frac{\epsilon \dot E}{ \dot M_{\rm hot}} \right)^{1/2}, 
\ee
where $\dot E$ is the total energy input from supernovae per unit time, $\epsilon$ is the fraction of this energy that is deposited into the hot medium, $\dot M_{\rm hot}$ is the total mass ejected from the galaxy  per unit time. 

We can solve eqs.\ (\ref{eq:mass})-(\ref{eq:energy}) outside the driving radius by simplifying the mass and momentum equations, using the fact that $\rho v r^2$ is a constant at these radii.
Defining  the kinetic and internal energy per unit mass as  $E_k = v^2/2$, and $I = (\gamma -1) c_s^2$, this gives
\ba
\frac{dE_k}{dt} &=& \left[\frac{1}{\gamma-1} - \frac{I}{2 E_k} \right]^{-1}  \left[ \frac{2 v I}{r} - \dot Q \right], \label{eq:Ek} \\  \qquad 
\frac{d I}{dt} &=&Ê\dot Q -\frac{dE_k}{dt},  \label{eq:I}
\ea
where $\dot Q \equiv - \Lambda \, n_b^2/\rho$ is the rate of radiative energy loss per unit mass, $n_b$ is the  number density of baryons (protons + neutrons) and $\Lambda$ is the cooling - heating rate as discussed in more detail below.  These equations can be used to track $v$ and $r$ as a function of time from the edge of the driving region for any choice of $\dot Q.$   The mean density of the medium as a function of time is then given by 
\be
\rho = \rho_\star (v/v_\star)^{-2} (r/r_\star)^{-1},
\ee   
where $v_\star$ and $\rho_\star$ are the velocity and density of the medium at the driving radius.
This approach allows the code to work in a comoving volume that travels outward with the medium, conserving mass.   I have modified the cosmology routines in the FLASH code to allow for such a  calculation, integrating eqs.\ (\ref{eq:Ek}) and (\ref{eq:I}) forward using a fourth-order Runge-Kutta method, and computing the expansion factor $a(t) = \left[\rho(t)/\rho_0 \right]^{-1/3}.$  Note that unlike usual cosmological calculations in which $a(t)=1$ at the end of the calculation, $a(t)=1$ at the onset of the calculation when $r=r_\star,$ and is greater than 1 at all subsequent times.

\subsection{Atomic, Ionic, and Compton Cooling and Heating}

In order to track atomic/ionic cooling  processes in the presence of significant photoionization from the central starburst, I followed the approach in \cite{2016arXiv160902561F}  and made use of the  cooling/heating functions developed in \citet{Gnedin12} \citep[see also][]{Sazonov05}.  This requires a calculation of 
\be
P_j \equiv c \int \sigma_j(\nu) n_\nu d \nu,
\label{eq:Pj}
\ee
where $n_\nu$ is the radiation field expressed as a number density of photons at a frequency $\nu$ and  $\sigma_j$ is the dissociation/photoionization cross section for $j \in$ $\{$H$_2$, HI, HeI, CIV$\}$.  For the starburst spectrum, I took a case in which the galaxy has been forming solar metallicity stars at a constant rate as modeled by  \citet{2003A&A...397..527S}.  Defining
\be
P_j \equiv \tilde P_j \frac{SFR}{M_\odot \, {\rm yr}^{-1}} \left(\frac{r}{300 \, {\rm pc}}\right)^{-2}, 
\ee
where SFR is the star formation rate. Making use of the analytic fits provided in \citet{Verner96}, 
this gives $\tilde P_j$ of $1.03 \times 10^{-7}$ s$^{-1}$, $5.28 \times 10^{-8}$ s$^{-1}$, $2.05 \times 10^{-8}$ s$^{-1}$ and 0  for Lyman-Werner dissociation, and HI, HeI, and CVI ionization, respectively.

In the left panel of Figure \ref{Fig:cool}, I show the cooling and heating rate normalized by the mean baryonic number density, $n_B^2,$ computed for a range of number densities from $0.01$ to $100$ cm$^{-3}$ for a medium at a distance of 300 parsecs from a starburst with $\dot M_{\rm hot}  = 10 M_\odot \, {\rm yr}^{-1}$.  At the very highest densities, recombination times are  short and the cooling rate above $10^4$ K is similar to the case without a background, although heating by photoionization is important below $10^4$K.    At 1 cm$^{-3}$, heating is boosted only slightly, but  the lowest temperature peak of the cooling function is significantly lessened, due primarily to hydrogen remaining much more ionized than would be the case in the absence of a background.    Finally, at the lowest densities, the mass fractions of neutral hydrogen, neutral helium, and other species with low-ionization potentials are extremely small, leading to a large decrease in the cooling rate at temperatures below $\approx 3 \times 10^{5}$K.  Together these effects shift the point at which cooling and heating are balanced from $\approx 10^4$K for gas with  $\rho/\mu m_p \gsim$ 10 cm$^{-3}$ to  $\approx 3 \times 10^4$K for gas with  $\rho/\mu m_p  \lsim$ 0.1 cm$^{-3}.$
 
In addition to atomic/ionic cooling as tabulated in these models, the presence of radiation from the starburst also leads to additional gas cooling and heating through Compton scattering with free elections.
To estimate this contribution, I modified the overall cooling/heating rate of the gas by a factor of
\be
H_C= \frac{\sigma_T F}{m_e c^2}  (\langle h\nu\rangle -4 k_B T) n_e =\Gamma_C \Delta\epsilon \,n_e, 
\label{eq34a} 
\ee
where  $F= L/(4 \pi r^2)$, $\langle h\nu \rangle = L^{-1} \int h\nu L_\nu d\nu$ = 9.14 eV such that $T_C \equiv \langle h\nu \rangle/4 k_B =2.65 \times 10^4$K, $L =  5.55 \times 10^{43}$ erg s$^{-1}$ SFR/ ($M_\odot$ yr$^{-1} $) is the luminosity per unit star formation rate (SFR). Including this effect shifts the cooling and heating curves from their \citet{Gnedin12} values to those shown in the right panels of Figure \ref{Fig:cool}, for a case with star formation rate of 10 $M_\odot$ yr$^{-1}$ and a distance of 300 pc.  Because the energy losses from Compton scattering are proportional to $n_e$ rather than density squared, they have the largest impact at the lowest densities, and because these losses are proportional to $T-T_C$, these have the largest effect when this gas is the hottest.  Thus Compton cooling is larger than atomic/ionic cooling for all temperatures  above $T \approx 10^6$K at densities lower than $\rho/\mu m_p  \lsim$ 10 cm$^{-3} \,   (\dot M_{\rm hot} / 10 M_\odot {\rm yr}^{-1}) (r/300 {\rm pc})^{-2}.$

\begin{table*}[t]
\begin{centering}
\caption{Simulation Parameters}
\resizebox{\textwidth}{!}{%
\begin{tabular}{|lccrrlccccc|}
\hline
  Name & SFR  & $\dot M_{\rm hot}$               & $\epsilon$  & $v_\star$ \,\,      &    $v_{\rm final}$ & $T_\star$   \,\, & $n_\star$  & Atomic & Compton & Dust  \\
          &  $M_\odot$ yr$^{-1}$	      &  $M_\odot$ yr$^{-1}$	    &    & km s$^{-1}$ & km s$^{-1}$ & $10^6$ K  & cm$^{-3}$ & Cooling & Cooling & Cooling \\
\hline

M82\_10\_1 & 10   &  10 & 1.0 & 500       & 1000 &    18 & 4.7  & N & N  & N \\
M82\_10\_1A& 10   &  10 & 1.0 & 500      & 970 &    18 & 4.7  & Y & N  & N \\
M82\_10\_1AC & 10    & 10 & 1.0 & 500   & 920   &    18 & 4.7 & Y & Y& N \\
M82\_10\_1ACD& 10   &  10 & 1.0 & 500  & 900    &    18 & 4.7  & Y & Y  & Y \\
M82\_5\_1ACD & 10   &  5   & 1.0 & 710     & 1320   &    36 & 1.7 & Y & Y & Y \\
M82\_2\_1ACD  & 10   &  2   & 1.0 & 1120  & 2120    &   90  & 0.42 & Y & Y& Y \\
M82\_1\_1ACD & 10   &  1   & 1.0 & 1580   & 3020  &  180 & 0.15 & Y & N & Y \\
M82\_5\_0.5ACD & 10 & 5   & 0.5 & 500   & 910 &   18  & 2.4  & Y & Y & Y\\
M82\_2\_0.5ACD  & 10 & 2   & 0.5 & 790   & 1480 &   45  & 0.6 & Y & Y & Y\\
ULIRG\_50\_1.0ACD & 100 & 50   & 1 & 710  & 1220 &   36  & 17  & Y & Y & Y\\
ULIRG\_20\_1.0ACD  & 100 & 20   & 1 & 1120  & 1950  &   90  & 4.2 & Y & Y & Y\\
ULIRG\_50\_0.5ACD & 100 & 50   & 0.5 & 500   & 860 &   18  & 24  & Y & Y & Y\\
ULIRG\_20\_0.5ACD  & 100 & 20   & 0.5 & 790  & 1360 &   45  & 6.0 & Y & Y & Y\\
\hline
\end{tabular}}
\end{centering}
\\
\end{table*}

\subsection{Dust Cooling and Destruction}

As in  \cite{2016arXiv160902561F}, the simulations include the impact of dust, which is able to provide cooling through two mechanisms. In the first mechanism, ion-grain and electron-grain collisions transfer thermal energy to the dust,
which is subsequently radiated away in the infrared. The cooling rate due to this mechanism can be written as $C_d^{\rm rad} = n_g H_{\rm coll}$
\citep{1981ApJ...245..880D} and \citet{1981ApJ...248..138D}
where $n_g ={\cal D}_0 n (\mu m_p/m_{g,0}) $ is the number density of dust particles in the gas,
with $m_{g,0}$ the initial grain mass, calculated by approximating the grains as sphere of radius $a$
and constant density $\delta_g,$
\be
m_g = 4 \pi a^3 \delta_g/3.
\ee
The heating rate deposited by particles into the grain (in ergs) can be parametrized  as
\ba
H_{\rm coll} = \begin{cases} 
5.38\times 10^{-10} n_e a^2 T^{3/2} &  x^* >4.5\\
1.47\times 10^{-3} n_e a^{2.41} T^{0.88} & x^* > 1.5\\
6.48\times 10^{6} n_e a^3 & x^* \le 1.5,
\end{cases} 
\label{eq32a} 
\ea
where $x^* = 1.26\times 10^{11} a^{2/3}/T$ (Montier \& Giard 2004). 

The second mechanism acts when a strong UV field is present, leading to positively-charged grains.  In this case the thermal energy of electrons ejected from the grain surface is less than the energy of electrons recombining, leading to a net cooling rate that depends on the flux in the UV (``Habing'') band 6-13.6 eV, $G_0=F/F_H$, where the source flux is normalized to the Habing flux $= 1.6\times 10^{-3} \ergs$ cm$^{-2}$ \citep{1968BAN....19..421H}, such that 
\be
G_0 = 1650  \frac{\dot M}{M_\odot \, {\rm yr}^{-1}} \left(\frac{R}{300 \, {\rm pc}}\right)^{-2}.
\ee
Then, the cooling rate due to recombinations on dust for a solar metallicity gas with a Milky Way dust-to-gas ratio is
\be
C_d^{\rm rec} \approx 4.65\times 10^{-30} \, \frac{{\rm erg}}{ {\rm s} \, {\rm cm}^{3}} \, \left(\frac{{\cal D}}{{\cal D}_\odot}\right) T^{0.94} \left(\frac{G_0 \sqrt{T} }{n}\right)^{\delta} 
n  n_e, 
\label{eq31} 
\ee
with $\delta=0.74/T^{0.068}$. 

Dust is included as a separate fluid, following its  evolution  using a passive tracer field  that travels with the flow \citep{2016arXiv160902561F}.
In each cell, ${\cal D} = {\cal D}_0 $ initially and dust is modeled as  a single-grain population, whose pre-shock radius is $\langle a \rangle = 0.1 \mu$m.
As the grains encounter the hot medium, they suffer a  gradual erosion due to thermal sputtering, described by a simple fit to the \citet{Dwek96} results \citep[see also][]{Draine79}:
\be
\frac{da} {dt} = - A n T_6^{-1/4} e^{-BT_6^{-1/2}},
\label{eq33} 
\ee
where $n$ and $T_6$ are the  density and temperature (in units of $10^6K$), and  $(A, B)=(1.2\times 10^{-5} \mu$m yr$^{-1}, 3.85)$. Finally, the rate of decrease of dust mass of a single grain is given by
\be
\frac{dm_g} {dt} = 4\pi a^2 \delta_g\frac{da} {dt},
\label{eq34} 
\ee
where the grain density $\delta_g = 3$ g $\cc$.

\section{Results}

Having implemented these code modifications, I carried out a set of simulations that span a large range of star formation rates (SFR), mass outflow rates ($\dot M_{\rm hot}$), and energy input rates $\dot E,$ as summarized in Table 1. All cases adoped a driving region with a radius of $r_\star = 300$ pc,  a value consistent with the nearby starburst M82 \citep{1995A&A...293..703M,2009ApJ...697.2030S} and also used in \cite{Zhang2015}.  The initial position of the simulation region was taken to be at $r_\star$ and the simulations tracked material within a 512$^3$ cell comoving box,  $150$ parsecs on a side, such that the comoving resolution was $0.29$ parsec.   All simulations assumed that  10$^{51}$ ergs of energy were released per 100 solar masses of stars formed, and that a fraction $\epsilon$ of the energy went into the wind.  As described in \S2.1, these values then determined the initial radial velocity and temperature of the material, $v_\star$ and $T_\star$, and the subsequent evolution of the flow determines the final value $v_{\rm final}$, which is 2 $v_\star$ in the absence of cooling.
Each of these values are also shown in Table 1. All simulations also assumed a total opening angle of outflowing material of $\Omega = \pi$ steradians as observed in large statistical samples of low-redshift starbursts \citep{2012ApJ...760..127M}, and this angle in turn determined the initial density of the material, as also given in the table.  

As they are approximating an initially thermalized medium, 
the simulations were run with a uniform initial temperature, but a fluctuating initial density field given by a white noise spectrum down to the scale of $1.2$ comoving parsecs (4 cells), and the amplitude of the fluctuations dropping by 50\% at $0.58$ parsecs and by 25\% at $0.29$ parsecs. The fluctuations were normalized to have an rms of $n_\star/8$ on 4 parsec scales, such that 
\be
n({\bf x}) = n_\star +  \frac{3 n_\star}{8} \left[ \frac{ r_4 +  0.5 r_2 + 0.25 r_1}{1+0.5/8^{1/2} + 0.25/8} \right],
\label{eq:ICs}
\ee
where are $r_1$, $r_2$, and $r_4$ are random numbers selected from a uniform distribution between -1 and 1, which vary on scales of 1, 2, and 4 cells respectively.
                         
The simulations were of two types,  a set with a SFR of 10 $M_\odot$ yr$^{-1}$, similar to that of the nearby starburst M82 \citep{2003ApJ...599..193F} and second set with a SFR of 100 $M_\odot$ yr$^{-1}$ similar to what is observed in ULIRGs \citep{2005ApJ...631L..13D,2012A&A...546A..64P}.  For each type of simulation, I varied the mass and energy input into the wind, and in the M82 case I also varied the physical assumptions, helping to disentangle the role that the various physical processes described in \S2 have on the evolution of the outflow.   

\subsection{Global Evolution and the Impact of Dust and Cooling Processes}

\begin{figure*}[t]
\center{\includegraphics[width=160mm]{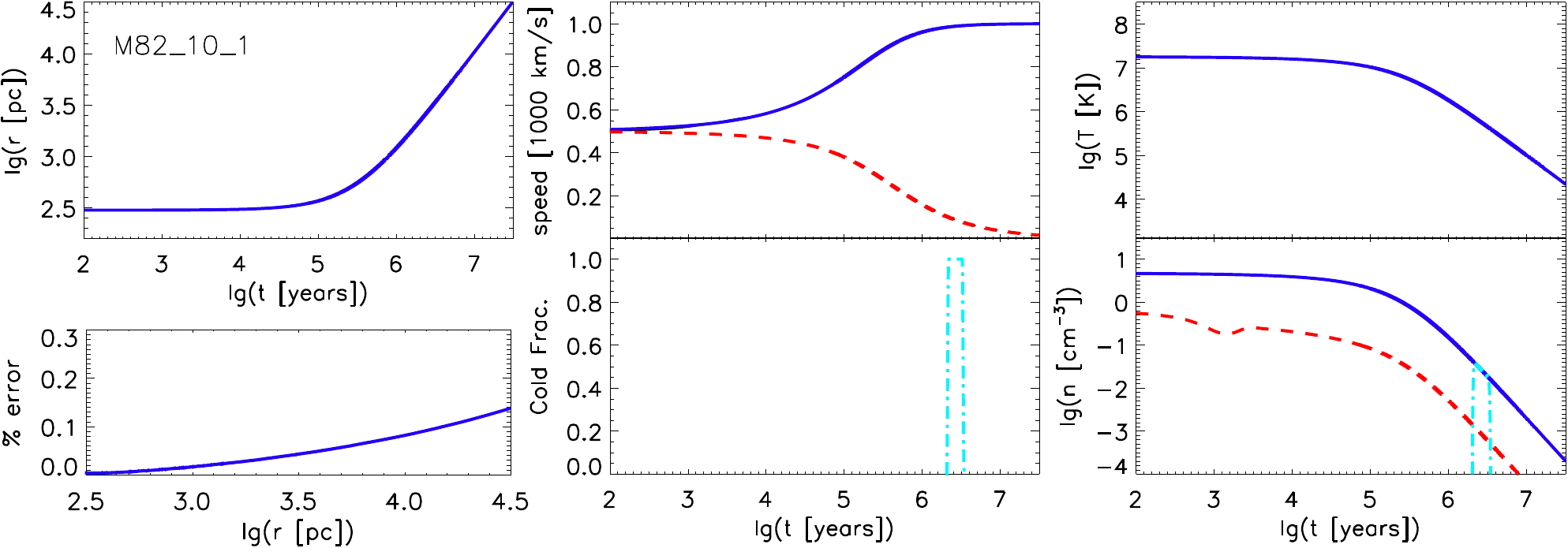}}
\caption{Evolution of the outflowing medium in the M82\_10\_1 case with $\dot M_{\rm hot}$ = 10 $M_\odot$ yr$^{-1}.$ and $v_\star = 500$ km s$^{-1}.$  {\em Top Left:} Distance between the center of the galaxy and the outflowing material.   {\em Top Center:} Velocity (blue solid curve)  versus sound-speed (red dashed  curve).  {\em Top Right:}  Average temperature.  {\em Bottom Left:} Fractional difference between the Mach number of the medium as calculated by FLASH and the analytic solution given in eq.\ (\ref{eq:ccMach}).   {\em Bottom Center:} Fraction of cold, $10^4 \, - \ 3 \times 10^4$ K, gas in the simulation.  {\em Bottom Right:} Average number density in the simulation (blue solid curve), rms average density fluctuations $(\left<n^2\right> - \left<n \right>^2)^{1/2}$ (red dashed curve), rms cold gas density $\left<n_{\rm cold}^2\right>^{1/2}$ (cyan dot-dashed curve) used as an estimate of emission from cold gas as described in \S 3.4. \vspace{0.18in}} 
\label{M82_10_1}
\end{figure*}

In the first simulation, denoted as M82\_10\_1, I took $\dot M_{\rm hot}$ = 10 $M_\odot$ yr$^{-1}$ and $\epsilon = 1.0$ (such that $\dot E = 10^{50}$ ergs / yr), I neglected all radiative cooling processes, and  assumed that the gas contained no dust. 
Although the outflow is a steady-state structure in this model, we can nevertheless plot the properties of the material as a function of time since passing though the driving radius. The top panels of Figure \ref{M82_10_1} show the evolution of the radial distance from the galaxy, radial velocity,   sound speed, and average temperature in this case.   Because $\dot Q = 0,$  an exact solution to eqs.\ (\ref{eq:mass}) - (\ref{eq:energy}) can be obtained as described in \cite{1985Natur.317...44C}.  In the region outside of the driving radius, this is 
\ba
\frac{r}{r_\star} &  =  & M^{1/(\gamma-1)} \left( \frac{\gamma-1+2 M^{-2}}{\gamma+1} \right)^{\left[\frac{\gamma+1}{4(\gamma-1)}\right]} \nonumber \\
& = & M^{3/2} \left( \frac{1+3 M^{-2}}{4} \right),
\label{eq:ccMach}
\ea
where the Mach number, $M \equiv v/c_s,$ and the second equality holds when $\gamma=5/3.$ The lower left panel of Figure \ref{M82_10_1} shows the fractional difference between this expression and the Mach number in the simulations. Here we see that the numerical and analytical results agree closely, differing by less than $0.2\%$ as the gas moves from as distance of 300 pc to 30 kpc from the host.  

The central panel of the lower row of Figure  \ref{M82_10_1} shows the fraction of the gas in the simulation with temperatures between $10^4$ K and  $3 \times 10^4$ K, a rough measure of the gas that will be visible though measurements of low-ionization state ions.   As cooling within the simulation is uniform and caused by adiabatic expansion, this leads to 100\% of the simulation moving into the cold state at a time set by the initial conditions and eq.\ (\ref{eq:ccMach}).  

Finally, the right panel of Figure  \ref{M82_10_1} shows the evolution of the average density in the simulation, the rms fluctuations about the mean density, $(\left<n^2\right> - \left<n \right>^2)^{1/2}$, and the rms cold gas density, $\left<n_{\rm cold}^2\right>^{1/2}$.  While the overall average density is determined directly by $v$ and $r$ as $n \propto (v r^2)^{-1},$ the rms density fluctuations give a measure of the inhomogeneity of the medium as a function of time, which must be solved for numerically.  The evolution of the rms fluctuations shows some early rearrangements, followed by a graduate decline in $(\left<n^2\right> \left<n \right>^{-2} -1 )^{1/2}$  as pressure differences smooth inhomogeneities as described in more detail below.   The decrease in $(\left<n^2\right> \left<n \right>^{-2} -1 )^{1/2}$  continues for roughly $\approx 10^6$ years, until the simulation enters the phase of high Mach number expansion, after which inhomogeneities stay largely constant.

This freeze-out of structure at late times can be understood from the fact that pressure differences can only smooth out structures whose sound speed is greater than their rate of expansion.  The overall expansion factor is $a(t) = [n(t)/n_\star]^{-1/3} = [v r^2/(v_\star r_\star^2)]^{1/3},$ such that at large radii, $a(t) \approx (2 r/r_\star)^{2/3}.$ This means, at large radii, the condition that the expansion rate of a clump of physical size $x_{\rm clump}$ be less than the sound speed of the medium gives
 \be
v_{\rm clump}  = x_{\rm clump} \frac{da}{dt} \approx x_{\rm clump} \frac{2}{3} \frac{v}{r} a < c_s,
\ee
which in comoving coordinates,  $x_{\rm clump,com} \equiv x_{\rm clump} \, a^{-1}$ gives
\be
x_{\rm clump,com} \lsim \frac{3 r}{2 M a^2} \approx r_\star \left[  \frac{1}{M}  \frac{3}{2^{7/3}} \frac{r^{1/3}_\star}{r^{1/3}} \right] \approx \frac{r_\star^2}{4 r},
\label{eq:clump}
\ee
where in the final expression I have also assumed that cooling is minimal such that eq.\ (\ref{eq:ccMach}) holds approximately and $a \approx 0.6 M.$  Thus the drop in the sound speed locks in density fluctuation into the medium when $r/r_\star$ becomes large, an issue we will return to in \S3.3.

Finally, as the density fluctuations never become large in this simulation, the rms cold gas density $\left<n_{\rm cold}^2\right>^{1/2}$ is very close to the average density during the period in which the cold fraction is one.  In this adiabatic simulation, this occurs when the mean density has dropped significantly, meaning that emission from low-ionization state ions is likely to be negligible in this model, as discussed in more detail in \S3.4.

\begin{figure*}
\center{\includegraphics[width=135mm]{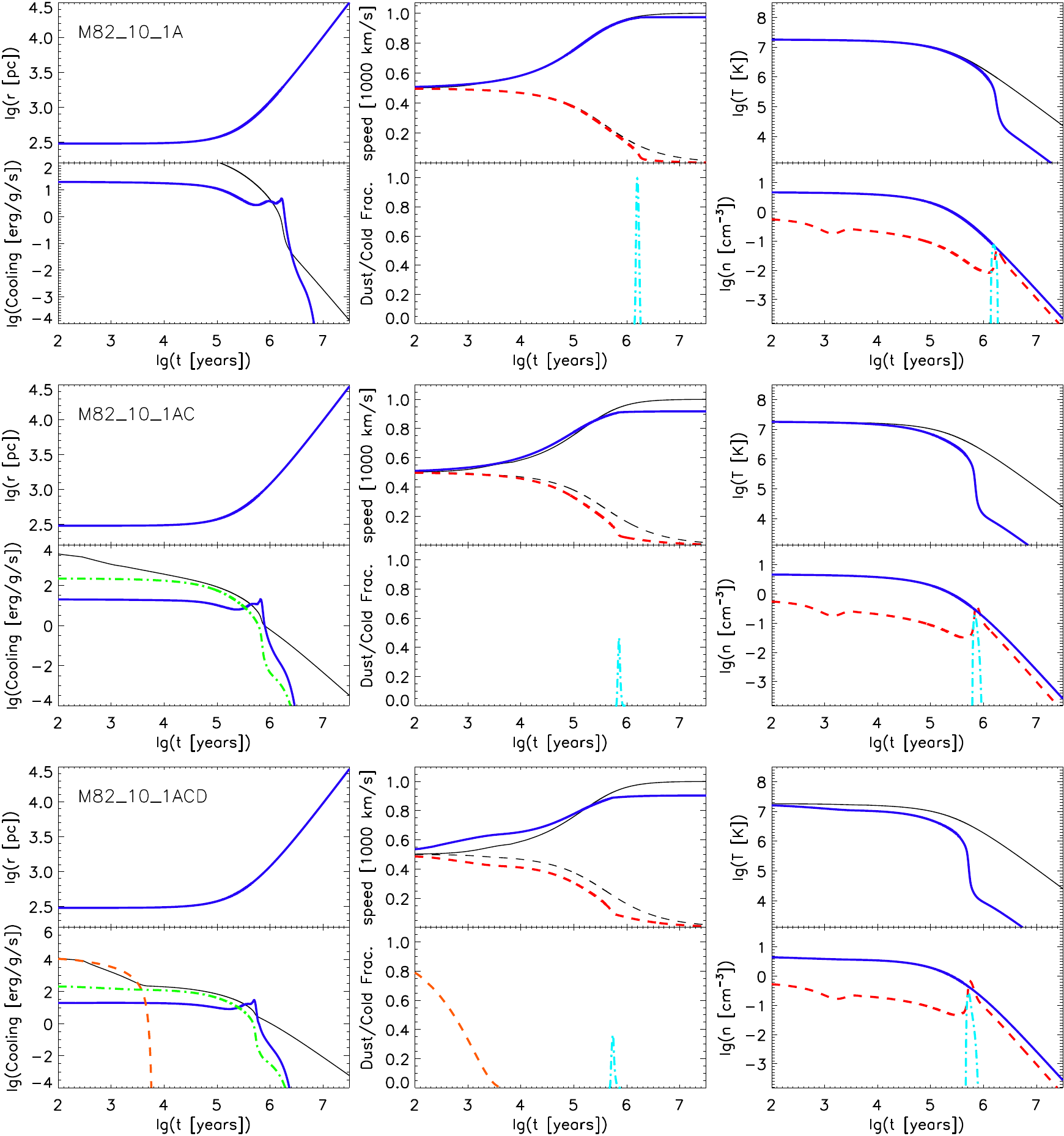}}
\caption{Evolution of global quantities in  $\dot M_{\rm hot}$ = 10 $M_\odot$ yr$^{-1},$ $v_\star = 500$ km s$^{-1}$ simulations including atomic/ionic cooling (case M82\_10\_1A, top two rows), atomic/ionic cooling and Compton cooling (case M82\_10\_1A, central two rows), and atomic/ionic  cooling, Compton cooling, and dust cooling and destruction (case M82\_10\_1ACD, bottom two rows).   In each set of rows the top left panel shows the evolution of the radius of the outflowing material, the top center panel shows the radial velocity (thick blue solid curve) and sound-speed (thick red dashed  curve) versus their values in the adiabatic case (thin black curves), and the top right panel shows the average temperature (thick blue solid curve) versus its value in the adiabatic case (thin black curve).   The bottom left panel in each set of rows shows the atomic/ionic (thick blue solid curve), Compton (green dot-dashed urve), and dust (red dashed curve) cooling rates, as compared to the adiabatic cooling rate (thin black curve), given by eq.\ (\ref{eq:adiacool}).  Finally, the bottom central panel in each set of rows shows the cold gas fraction (cyan dot-dashed curve) and the fraction of the initial dust (red dashed), and the bottom right panel shows the average number density (blue solid curve), rms average density fluctuations (red dashed curve), and rms average cold gas density (cyan dot-dashed curve). \vspace{0.18in}} 
\label{M82_physics}
\end{figure*}

Figure \ref{M82_physics} examines the impact of additional cooling processes on the evolution of an M82 type outflow with  $\dot M_{\rm hot}$ = 10 $M_\odot$ yr$^{-1}$ and $\dot E = 10^{50}$ ergs / yr.  The figure shows results for three different cases: case M82\_10\_1A including atomic/ionic cooling assuming solar metallicity, case M82\_10\_1AC, including atomic/ionic and Compton cooling, and case M82\_10\_1ACD, including atomic/ionic  cooling, Compton cooling, and dust cooling and destruction  assuming an initial Milky Way  dust-to-gas ratio.   For comparison, in these plots I also compute the rate of cooling due to adiabatic expansion:
\be
\left( \frac{dI}{dt} \right)_{\rm adiabatic} = I (\gamma-1) \frac{d {\rm ln} n}{dt}.
\label{eq:adiacool}
\ee

In the M82\_10\_1A, case atomic/ionic cooling overtakes the adiabatic cooling rate at $\approx 10^6$ years, when the temperature has dropped to $\approx 10^6$K and emission processes are very efficient.   At this point, the sound speed and temperature drop rapidly, causing a decrease in pressure that leads to a radial acceleration that is larger than in the adiabatic case.   On the other hand, this same pressure decrease means that at very large radii the radial velocity is always less than in the adiabatic case, leading to a final velocity of 970 km s$^{-1}$ as opposed to 1000 km s$^{-1}.$

A more dramatic effect of atomic/ionic cooling is its impact on the evolution of density fluctuations.   Because cooling occurs first in the densest regions, the pressure within them drops significantly with respect to their surroundings, leading to a rapid increase in $(\left<n^2\right> \left<n \right>^{-2} -1 )^{1/2}$.  Thus radiative cooling not only causes the gas to reach  $10^4 \, - \ 3 \times 10^4$K temperatures at higher densities, but also amplifies densities inhomogeneities, increasing the observability of the medium.

Moving to the  M82\_10\_1AC  case that includes Compton cooling, we uncover even larger changes with respect to the adiabatic case.   In fact, for the first 300,000 years of evolution, cooling by this process is over an order of magnitude more efficient than atomic/ionic cooling, and almost as efficient as adiabatic cooling itself.  This causes the temperature to be significantly lower at very early times and leads to a short phase in which $v$ exceeds the adiabatic value, but it ultimately causes the final velocity to be  920 km s$^{-1}$ as compared to 1000 km s$^{-1}$.   Because Compton cooling depends on the scattering of electrons with photons, the energy loss per unit mass is independent of density, and unlike atomic/ionic cooling, it does not lead to the growth of inhomogeneities. On the other hand, it allows the gas to cool into the $10^4 \,-\ 3 \times 10^4$K range at smaller radii and at higher densities.

Finally, the lower two rows of Figure \ref{M82_physics} show the results from the M82\_10\_1ACD simulation that includes dust cooling and destruction.  While dust cooling is extremely efficient at the high densities that occur when $r \approx r_\star$ in this model, these high densities also lead to rapid dust sputtering from collisions between dust particles and the $10^7$K medium.  Thus, at least for this choice of parameters, dust makes a difference only at the very earliest times, having an overall impact that is  similar to slightly reducing $\epsilon,$ the fraction of the available supernova energy added into the hot medium.

\begin{figure*}
\center{\includegraphics[width=160mm]{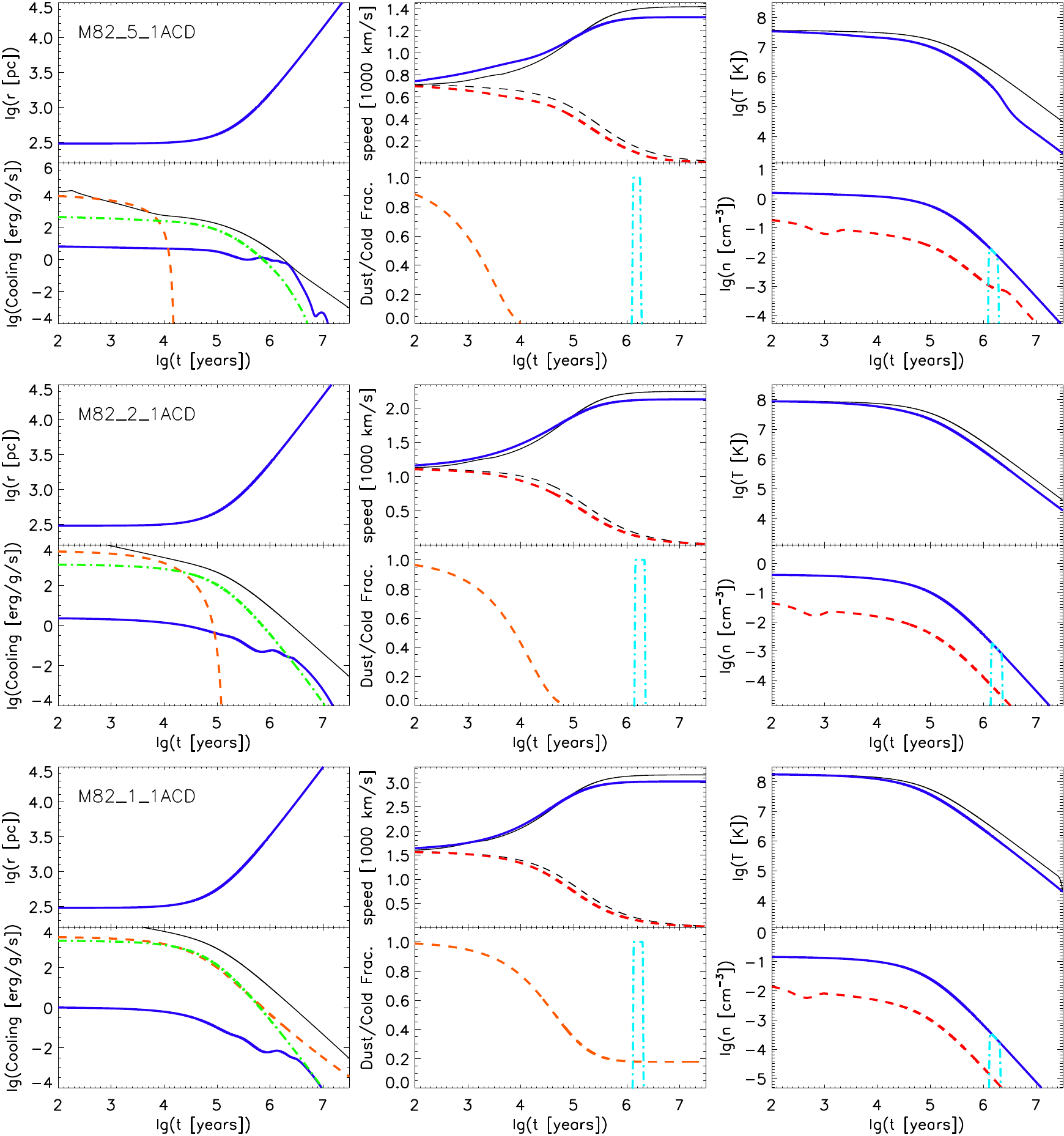}}
\caption{Evolution of global quantities in M82-type simulations including  dust cooling and destruction in addition to atomic/ionic cooling and Compton cooling. The same three physical parameters are shown as in Fig.\  \ref{M82_dust}:  $\dot M_{\rm hot}$ = 5 $M_\odot$ yr$^{-1},$ and $v_\star = 710$ km s$^{-1},$  (case M82\_5\_1ACD top two rows), $\dot M_{\rm hot}$ = 2 $M_\odot$ yr$^{-1},$ and $v_\star = 1120$ km s$^{-1},$  (case M82\_2\_1ACD central two rows), and $\dot M_{\rm hot}$ = 1 $M_\odot$ yr$^{-1},$ and $v_\star = 1580$ km s$^{-1},$  (case M82\_1\_1ACD bottom two rows).  In each set of rows the panels and curves are as in Fig.\ \ref{M82_physics}. \vspace{0.18in}}
\label{M82_dust}
\end{figure*}

\subsection{Effect of Varying Outflow Properties}

Next we turn our attention to the relationship between the initial wind properties and the subsequent evolution of the outflowing gas.   In Figure \ref{M82_dust},  as in the previous section, we consider three M82-like cases in which $r_\star$ = 300 parsec, the SFR = 10 $M_\odot$ yr$^{-1},$ and $\dot E = 10^{50}$ ergs / yr, but now we take $\dot M_{\rm hot}$ = 5 $M_\odot$ yr$^{-1}$ (case M82\_5\_1AC with $v_\star = 710$ km s$^{-1}$), $\dot M_{\rm hot}$ = 2 $M_\odot$ yr$^{-1}$ (case M82\_2\_1AC with $v_\star = 1120$ km s$^{-1}$) and $\dot M_{\rm hot}$ = 1 $M_\odot$ yr$^{-1}$ (case M82\_1\_1AC with $v_\star = 1580$ km s$^{-1}$).  Note that the last of these cases corresponds to pure supernova ejecta with no energy losses and no additional entrained material.  All models include atomic/ionic cooling, and cooling by Compton scattering and dust emission.    Decreasing the mass outflow rate while leaving the energy input rate unchanged has several important effects. First it leads to higher temperatures and pressures, which in turn boost the radial velocity of the flow.   This means that the density of the medium decreases due both to lower mass fluxes, and also due to the material is moving outward more rapidly.  As atomic/ionic cooling and dust cooling per unit mass are proportional to $n$, this means that lower $\dot M_{\rm hot}$ cases not only start with more internal energy per particle, but this energy is more difficult to radiate away.

\begin{figure*}[t]
\center{\includegraphics[width=160mm]{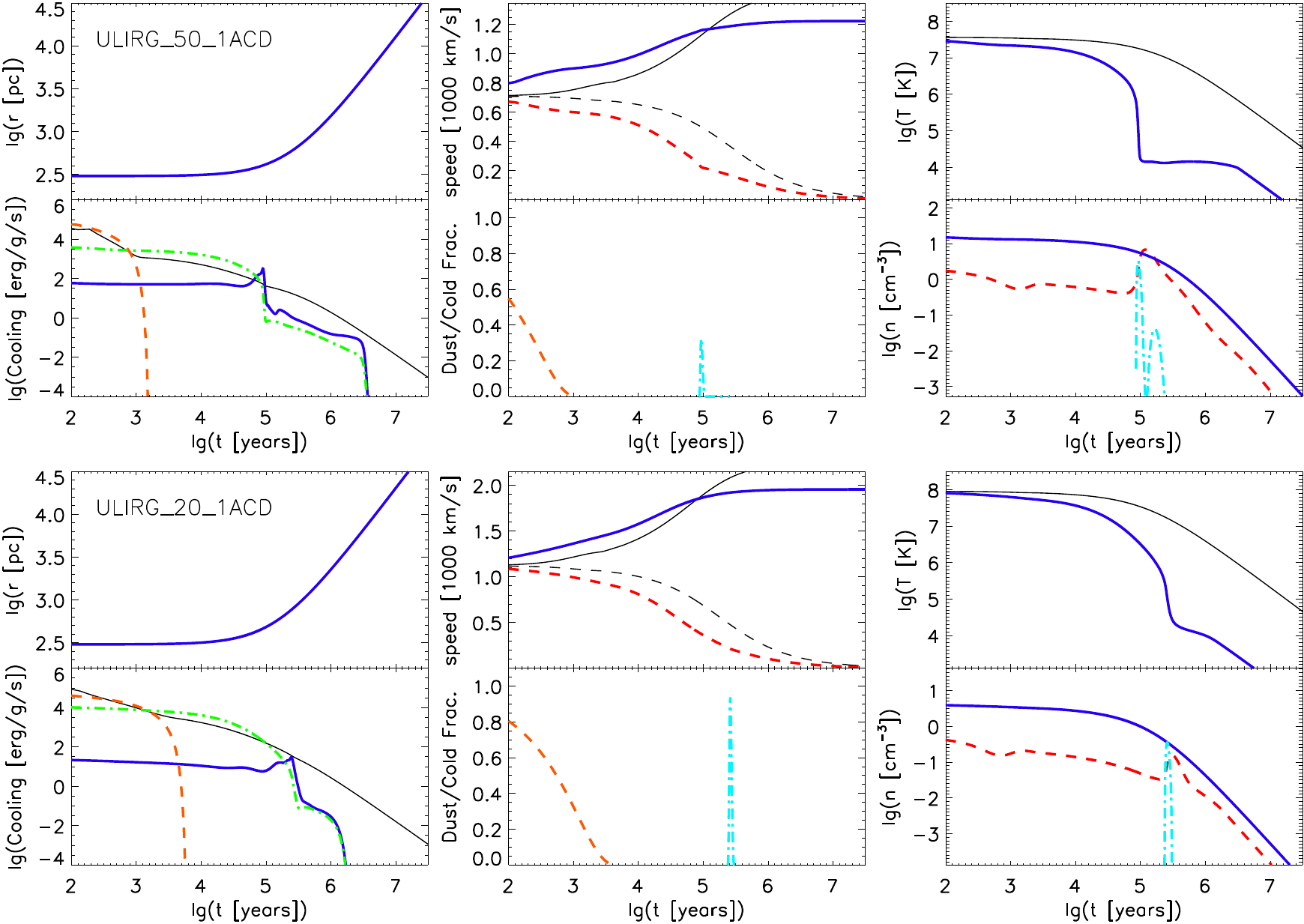}}
\caption{Evolution of global quantities in ULIRG-type simulations including  atomic/ionic cooling, Compton cooling, and dust cooling an destruction. The model parameters are:  $\dot M_{\rm hot}$ = 50 $M_\odot$ yr$^{-1},$ and $v_\star = 710$ km s$^{-1},$  (case ULIRG\_5\_1ACD, top two rows), $\dot M_{\rm hot}$ = 20 $M_\odot$ yr$^{-1},$ and $v_\star = 1120$ km s$^{-1},$  (case ULIRG\_2\_1ACD, bottom two rows). In each set of rows the panels and curves are as in Figs.\ \ref{M82_physics}  and \ref{M82_dust}. \vspace{0.18in}}
\label{ULRIG_dust1}
\end{figure*}

On the other hand, as the Compton cooling rate per particle is proportional to the gas temperature and independent of the gas density this process increases in importance in lower $\dot M_{\rm hot}$ cases.  Thus in the M82\_5\_1ACD case, Compton cooling dominates over atomic/ionic cooling for roughly the first $10^6$ years of evolution, as opposed to $\approx 3 \times 10^5$ years in the  M82\_10\_1ACD case.  While dust cooling is rapid, it is limited to very early times, much as was found in the  M82\_10\_1ACD case.   All together, these processes are less important than adiabatic expansion.  Thus cooling has only a moderate impact in this simulation, decreasing the final velocity of the flow by $\approx 100$ km/s, and slightly increasing the  inhomogeneity of the medium.

Radiative cooling is even less important in the $\dot M_{\rm hot}$ = 2 $M_\odot$ yr$^{-1}$ case shown in the central panels of Figure  \ref{M82_dust}.  In this case atomic/ionic cooling is always at least an order of magnitude smaller than adiabatic cooling, causing it to have little impact on the evolution of the medium or the evolution of the density structure.  Furthermore, while Compton cooling is significantly higher in the initially hot medium, it still falls far short of the adiabatic cooling rate.   While dust cooling is also much smaller than adiabatic cooling, dust sputtering proceeds slowly,  due to the lower densities, which leads to appreciable dust densities up to $\approx 10^5$ years after the material has moved out of the driving region. This is late enough that the material has undergone significant acceleration, and is well into the supersonic regime.    In this case, no stage of rapid cooling is seen whatsoever, and the difference between the final velocity (2120 km/s) and the final velocity in the adiabatic case (2240 km/s) is less than 5\%.

In the  M82\_1\_1ACD case with $\dot M_{\rm hot}$ =  1 $M_\odot$ yr$^{-1},$ Compton, atomic/ionic, and dust cooling are to  too week to lead to observable effects, and the reduction in the final velocity is again only a few percent. Interestingly however,  gas densities in this model have dropped to the point that sputtering is no longer able to destroy the dust before it leaves the galaxy. This means that this case is able to deposit dust into the surrounding circumgalactic medium directly through the hot medium, without requiring the any cold clouds to be entrained by the hot gas. This is the hottest and most rapidly-moving of all the simulated cases, but it is also the only one in which dust is not completely destroyed within the hot wind.  This suggests that, perhaps counterintuatively, starbursting galaxies whose winds entrain the least amount of material may nevertheless be the most efficient at depositing the dust observed at distances of 20 kpc to several Mpc outside of galaxies \citep{2010MNRAS.405.1025M,2012ApJ...754..116M}.

Table 1 also include describes models including energy losses within the driving region, in which the fraction of supernovae energy deposited in the outflowing medium was taken to be 50\%, $\epsilon = 0.5.$   In the M82\_5\_0.5ACD case with $\dot M_{\rm hot}$ = 5 $M_\odot$ yr$^{-1},$ this reduces the initial velocity to $v_\star = 500$ km s$^{-1},$ from 710  km s$^{-1}$ and in the M82\_2\_0.5ACD case with $\dot M_{\rm hot}$ = 2 $M_\odot$ yr$^{-1},$ it reduces the initial velocity to  $v_\star = 790$ km s$^{-1},$ from 1120 km s$^{-1}.$   It also reduces the initial temperature of the winds by a factor of $2$ with respect to their $\epsilon = 1$ values.  Such slower, colder winds, are naturally much more susceptible to cooling. In the $\dot M_{\rm hot}$ = 5 $M_\odot$ yr$^{-1}$ case, this leads to an $\approx 9\%$ reduction in the final velocity and the rapid production of heterogeneous, cold gas within $\approx 1$ kpc of the galaxy.  In the 2 $M_\odot$ case, cooling processes are more modest, such that the final velocity is reduced by  $\approx 6\%$, and no phase of rapid structure growth occurs. In both cases, the densities are high enough that dust is rapidly destroyed by sputtering.

\begin{figure*}
\center{\includegraphics[width=160mm]{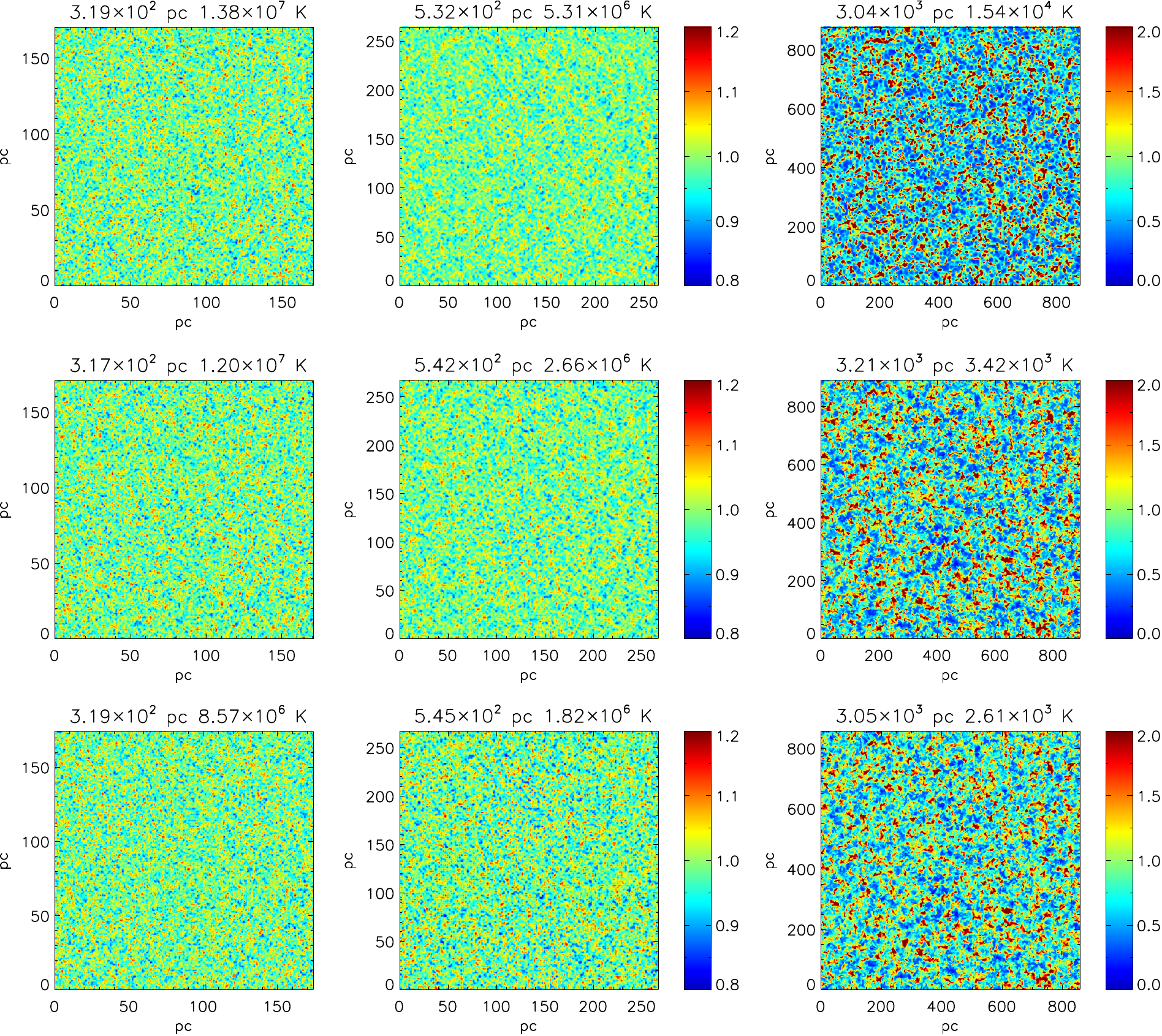}}
\caption{Normalized density distribution, $n/\left< n \right>,$ on slices taken from M82 type simulations with SFR = $\dot M_{\rm hot}$ = 10 $M_\odot$ yr$^{-1}$ and $\epsilon =1.$  {\em Top row:} Results from case M82\_10\_1A, which includes only atomic/ionic cooling.  The times are $3.0 \times 10^4$, $2.9\times10^5$, and $2.9\times10^6$ years. {\em Center row:} results  from case M82\_10\_1AC, which also includes Compton cooling. Here the times are $2.7\times10^4$ years (left), $3.0\times10^5$ years (center), and $3.2\times10^6$ years (right). {\em Bottom row:} Results from case M82\_10\_1ACD, which also includes dust cooling and destruction.   The times are $2.8\times10^4$, $3.0\times10^5$, and $3.0\times10^6$ years.    Each panel is labeled with its corresponding value of  the outflow radius and the mean temperature, and the axes are labeled in physical (not comoving) coordinates.  Note that the times differ slightly between the cases because the time steps taken by the different simulations do not correspond exactly. \vspace{0.18in}}
\label{M82_10_1_slice}
\end{figure*}

\begin{figure*}
\center{\includegraphics[width=160mm]{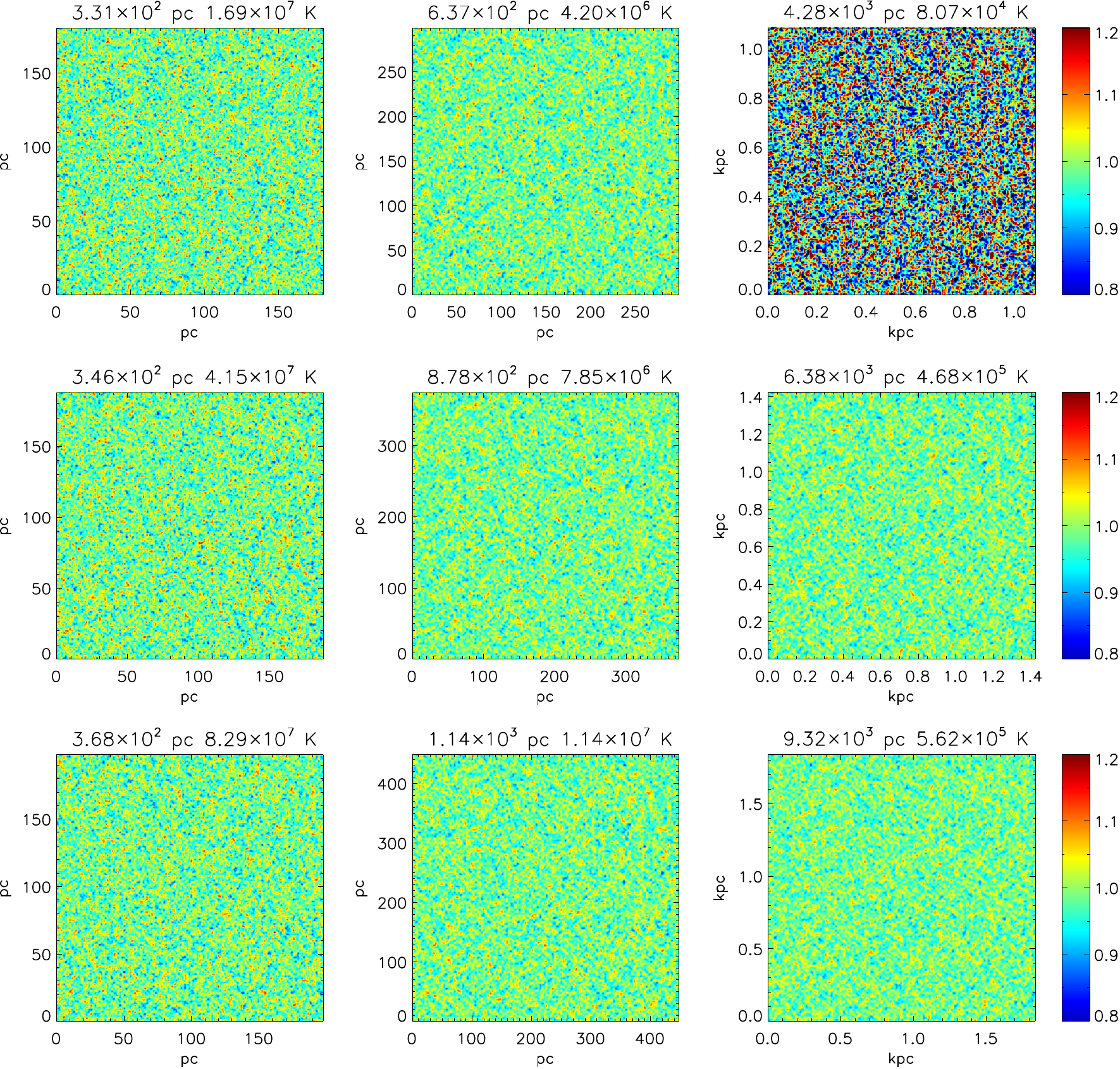}}
\caption{Normalized density distribution, $n/\left< n \right>,$ on slices taken from M82-type simulations including atomic/ionic, Compton, and dust cooling, and with mass outflow rates reduced from the fiducial value of 10 $M_\odot$ yr$^{-1}$.  {\em Top row:} Results from case M82\_5\_1ACD, with a mass outflow rate of 5 $M_\odot$ yr$^{-1}.$ The times are  $3.2\times10^4$, $2.9\times10^5$, and $3.0\times10^6$ years.  {\em Center row:} Results from case M82\_2\_1ACD,  with a mass outflow rate of 2 $M_\odot$ yr$^{-1}.$  The times are  $3.0\times10^4$, $3.0\times10^5$, and $2.9\times10^6$  years.  {\em Bottom row:} Results from case M82\_1\_1ACD, with a mass outflow rate of 1 $M_\odot$ yr$^{-1}.$  The times are  $3.1\times10^4$, $3.0\times10^5$, and $3.0\times10^6$ years. Each panel is labeled with its corresponding value of the outflow radius and the mean temperature, and the axes are labeled in physical (not comoving) coordinates. \vspace{0.18in}}
\label{M82_dust_slice}
\end{figure*}

Figure \ref{ULRIG_dust1} shows the results of simulations of a more extreme case of a ULIRG type galaxy with a SFR of $\dot M$ = 100 $M_\odot$ yr$^{-1}$ and  $\epsilon = 1$ such that the energy input is $10^{53}$ ergs yr$^{-1}$.  Such galaxies have similar sizes as M82 \citep[e.g.][]{2000AJ....119..509S}, and so for ease of comparison I take $r_\star = 300$ pc.  I also consider mass outflow rates of  50 $M_\odot$ yr$^{-1}$ and  20 $M_\odot$ yr$^{-1}$ such that $M_\odot$ yr$^{-1}$/SFR is 0.5 and 0.2 as in the M82 cases above.   Thus the initial conditions in the ULIRG\_50\_1ACD case are the same as in the M82\_5\_1ACD case except the number density of the gas and the photon flux from the galaxy are both shifted upward by a factor of 10. This in turn increases the importance of atomic/ionic, Compton, and dust cooling, boosting them well above adiabatic cooling. The result is widespread cooling and growth of inhomogeneities within $10^5$ years, at a distance just outside the driving radius.   Thus, unlike the M82\_5\_1ACD case, the ULIRG\_50\_1ACD case is so radiatively efficient that it is $10^4$K throughout the full evolution outside the driving region.  In the 20 $M_\odot$ yr$^{-1}$ case, cooling processes are slightly reduced, but again the gas cooling occurs very near to the galaxy.   Here $10^4$K temperatures are reached within $3 \times 10^5$ years and $r \approx 600$ pc.

Table 1 also describes ULIRG models with outflow rates of  50 $M_\odot$ yr$^{-1}$ and  20 $M_\odot$ yr$^{-1}$  in which the efficiency of energy input is  50\% ($\epsilon=0.5$). The higher densities and lower temperatures in these models boost the cooling efficiencies even further such that the final velocities are much lower than their ($2 v_\star$) adiabatic values and   $10^4$K temperatures are reached within $10^5$ years, corresponding to radii that are barely beyond the driving radius.   Furthermore,  although the star-formation in ULIRGs is dust-enshrouded,  the high densities and temperatures in the outflow mean that dust is destroyed almost immediately as it leaves the galaxies.  This suggests that the most likely state of outflowing material from the dustiest, most rapidly star-forming galaxies in the Universe is cold and dust-free.

\subsection{Growth of Structure}

Figure \ref{M82_10_1_slice} shows the density distribution  on slices taken from the M82 type simulations with various cooling processes included,  at representative times of  $\approx 3\times10^4,$ $\approx 3\times10^5$,  and $\approx 3\times10^6$ years.  Each  panel is labeled with its corresponding values of  the outflow radius and the mean temperature.  Between $3 \times 10^4$ and $3 \times 10^5$ years, cooling is dominated by adiabatic expansion and by Compton cooling (in the simulations in which it is included).    Because both these processes lead to uniform cooling per unit mass, they do not cause pressure differences, and thus the overall changes in the structures between these two times is minor.  

\begin{figure*}
\center{\includegraphics[width=160mm]{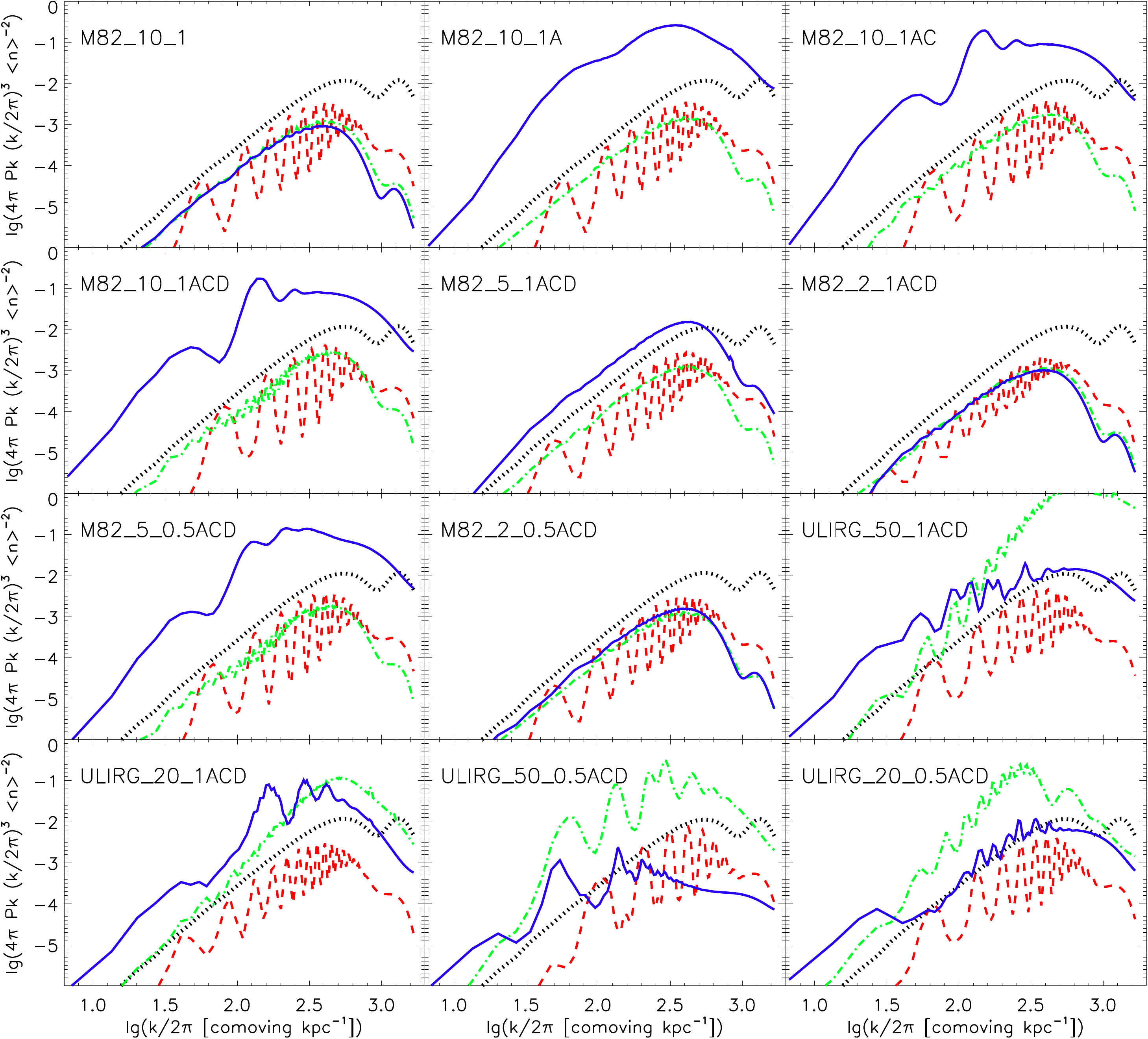}}
\caption{Density power spectra as a function of time in the simulations.  Each panel is labeled by the simulation from which it was taken, moving from the fiducial M82 case with various physical processes included, to M82 cases with lower mass outflow rates and $\epsilon$ values, to ULIRG cases with a variety of mass outflow rates and $\epsilon$ values.  In each panel, results are shown at $t=0$ (dotted black curves),  $t \approx 3\times10^4$ years (red dashed curves) $t \approx 3\times10^5$ years (green dot-dashed curves), and $t \approx 3\times10^6$ years (blue solid curves). \vspace{0.18in}}
\label{dspectra}
\end{figure*}

When the medium cools below $\approx 10^6$K however, the efficiency of atomic/ionic cooling increases dramatically.  Unlike expansion and Compton cooling, this process is density-dependent, leading to the most rapid cooling per unit mass in the overdense regions.  As these regions cool, the higher pressure gas around them compresses them further, leading to a sharp increase in the magnitudes of density fluctuations, as evidenced in the increase in the rms density fluctuations seen in Figure \ref{M82_physics}.  Thus by $3 \times 10^6$ years, the density fluctuations have increased from their values at $3 \times 10^5$ years by roughly a factor of 10.  This rapid growth of fluctuations is soon halted, however, as the exterior medium has a chance to cool and pressure contrasts become much more modest.  At this time the average temperature in the medium is $\lesssim 10^4$K, which corresponds to a sound speed of $\approx 10$ km/s.    As the medium is moving $\approx 1000$ km/s, its expansion is rapid with respect to these motions.   From eq.\ (\ref{eq:clump}) this means that all fluctuations with comoving sizes above $\approx 0.6 r_\star M^{-1} (r/r_\star)^{-1/3} \approx 2 \, {\rm pc} (r/r_\star)^{-1/3}$ are subsequently frozen into the medium.

Figure \ref{M82_dust_slice} shows the density distribution  on slices taken from M82 type simulations with mass outflow rates smaller than the fiducial value of 10 $M_\odot.$ Note that in this figure, the density fluctuations are shown on the same scale in all panels.  As in the  10 $M_\odot$ case, above there are only minor changes in the density distribution in the period between $3 \times 10^4$ and $3 \times 10^5$ years, when adiabatic and Compton cooling are dominant.  At later times, however, the evolution of the medium is strongly dependent on the mass  outflow rate.  The M82\_5\_1ACD case with  $\dot M_{\rm hot}$ = 5 $M_\odot$ yr$^{-1}$ shows a significant increase in the magnitude of density fluctuations at $3 \times 10^6$ years, which is caused by density-dependent atomic/ionic cooling.   However, because the mean density in the simulation is less than in the $\dot M_{\rm hot}$ = 10 $M_\odot$ yr$^{-1}$ case, the overall impact of this cooling is also less, increasing density fluctuations by a factor of $\approx 3$, rather than a factor of 10 as seen above.

In the 2 $M_\odot$ yr$^{-1}$ case, the impact of cooling from radiative processes in decreased even further, such that by the time the simulation moves into the $\lesssim10^6$ temperature regime, the outflow has expanded to the point that density fluctuations are already largely frozen into the medium.   Thus there is no particular increase in the rms fluctuations at $3 \times 10^6$ years seen in this model in Figure  \ref{M82_dust_slice}, which is also consistent with the overall evolution of $(\left<n^2\right> - \left<n \right>^2)^{1/2},$ as shown in Figure \ref{M82_dust}.  Similarly, in the  1 $M_\odot$ yr$^{-1}$ case, cooling from atomic/ionic processes is even less important, and there is no phase in which density contrasts are increased.

To quantify these effects further, I computed the density power spectra from the simulations as
\be
P(k) \equiv \left< \tilde n(k) \tilde n^*(k) \right>, 
\ee
where $n({\bf k}) \equiv \int d^3 x \, n({\bf x})  \, \exp(- i {\bf k \cdot x})$ and the wave number $k$ is equal to $2 \pi$ divided by the wavelength of the perturbation.
In this case, the total density fluctuations in units of the mean density are
\be
\frac{\left<n^2\right>}{\left<n \right>^2} -1 = \frac{4 \pi}{\left<n \right>^2} \int \left(\frac{k}{2 \pi} \right)^3 P(k) {\rm d \, ln}k.
\ee
Figure \ref{dspectra} shows the evolution of the structure in the simulations normalized as $4 \pi (k/2 \pi)^3 P(k) \left<n \right>^{-2},$ meaning that the maximum values of these curves will be roughly equal to the variance of the density fluctuations in units of the mean density.

In the adiabatic M82\_10\_1 case, which appears in the upper left panel of this figure, there are no processes in place to enhance density inhomogeneities, leading to a steady decline in the power spectrum over time.   This is most dramatic in the interval from $t=0$ to $t \approx 3 \times 10^4$ in which pressure differences in the  initial conditions act to rapidly smooth structure.  While this smoothing occurs on all scales, it is most significant at the shortest distances, corresponding to  $2 \pi/k \lesssim 2$ comoving parsecs on this plot.  Note also that the simplified initial conditions as given by eq.\ (\ref{eq:ICs}) cause some aliasing in the $3 \times 10^4$ power spectrum, leading to a saw-tooth pattern that would not be expected in a less-idealized configuration.    The pressure smoothing continues at later times, strongly suppressing structures below $\approx 3$ comoving parsecs by $3 \times 10^5$ years, but having a much more subtle impact at later times, as the medium expands to the point that the remaining homogeneities become frozen into the medium.

\begin{figure*}
\center{\includegraphics[width=170mm]{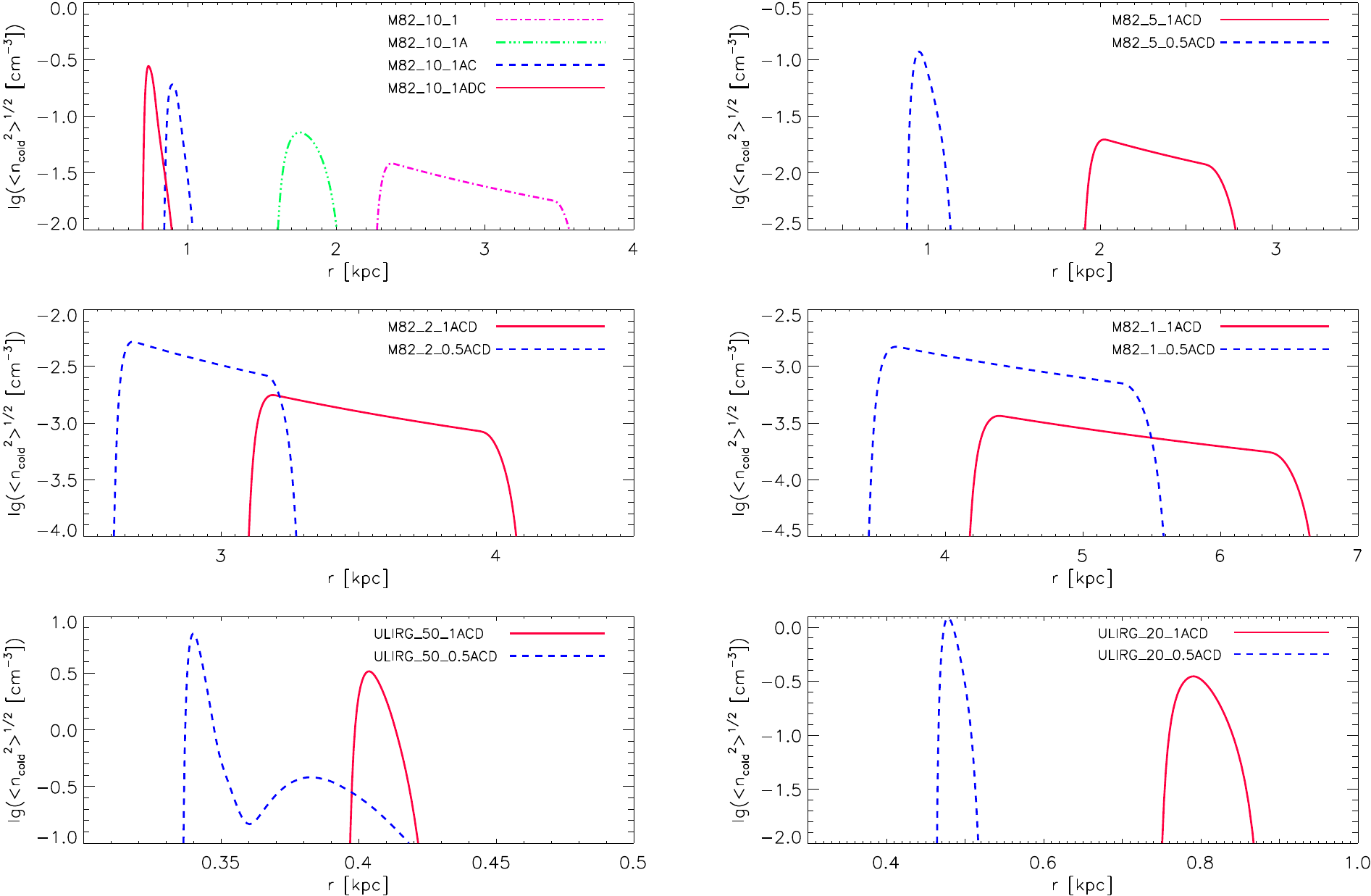}}
\caption{Rms cold gas density, $\left< n_{\rm cold}^2 \right>^{1/2},$ which provides an estimate of the emission from cold gas as a function of radius.   {\em Top Left:}  Radial profiles from fiducial M82 simulations with varying cooling physics: adiabatic (dot-dashed magenta curve), atomic/ionic cooling (triple-dotted dashed green curve), atomic/ionic + Compton cooling (dashed blue curve),  atomic/ionic, Compton, and dust cooling (solid red curve).  {\em Top Right:} Results from M82 simulations with a mass outflow rate of 5 $M_\odot$ yr$^{-1},$ and with $\epsilon = 1$ (solid red curve), and $\epsilon = 0.5$ (dashed blue curve). {\em Center Left:} Results from M82 simulations with a mass outflow rate of 2 $M_\odot$ yr$^{-1},$ and with $\epsilon = 1$ (solid red curve), and $\epsilon = 0.5$ (dashed blue curve). {\em Center Right:} Results from M82 simulations with a mass outflow rate of 1 $M_\odot$ yr$^{-1}, $ and with $\epsilon = 1$ (solid red curve), and $\epsilon = 0.5$ (dashed blue curve). {\em Bottom Left:} Results from ULIRG simulations with a mass outflow rate of 50 $M_\odot$ yr$^{-1}, $  and with $\epsilon = 1$ (solid red curve), and  and $\epsilon = 0.5$ (dashed blue curve).  {\em Bottom Right:} Results from ULIRG simulations with a mass outflow rate of 20 $M_\odot$ yr$^{-1}, $  and with $\epsilon = 1$ (solid red curve),  and $\epsilon = 0.5$ (dashed blue curve). \vspace{0.18in}}
\label{fig:coldr}
\end{figure*}

In the M82\_10\_1A case in which atomic and ionic cooling are included, the early evolution is almost identical to that seen in the M82\_10\_1 case.  Between $3 \times 10^5$ years and $3 \times 10^6$ years, however, as the outflow cools adiabatically to below $\approx 10^6$K, density-dependent radiative cooling becomes dominant.   This leads to the rapid growth of structure, boosting the power spectrum at all scales, and increasing the overall rms value of the fluctuations by a factor $\approx 10$,
as also seen in Figure \ref{M82_10_1_slice}.   

The inclusion of Compton and dust cooling, as shown in cases M82\_10\_1AC and M82\_10\_1ACD has only a minor impact of the overall evolution of the power spectra, but the impact of changing  the mass outflow rate is much more significant.  Thus the curves from the M82\_5\_1AC and M82\_2\_1ACD cases illustrate quantitatively how decreasing $\dot M_{\rm hot}$ from 10 $M_\odot$ yr$^{-1}$ to  5 $M_\odot$ yr$^{-1}$ leads to a muted period of structure growth through density-dependent cooling, while decreasing it to 2 $M_\odot$ yr$^{-1}$ leads to an overall evolution that is extremely similar to the adiabatic M82\_10\_1A case.    

Similarly, when the fraction of the supernova energy that is deposited into the hot medium is reduced to $\epsilon = 0.5,$ not only is the radial velocity of the outflow reduced, but the density of the medium goes up, increasing the efficiency of radiative cooling.   In the M82\_5\_0.5ACD case with 5 $M_\odot$ yr$^{-1}$ and $\epsilon=0.5$, this causes the phase of structure growth from density-dependent cooling to become much more significant, leading to an overall history similar to the M82\_10\_1ACD case.    On the other hand, when the mass outflow rate is 2 $M_\odot$ yr$^{-1},$  reducing $\epsilon$ from 1 to 0.5 does not increase the importance of radiative cooling to the point that it leads to significant structure growth.   Instead, the large difference in structure growth between these two similar models means that the observability of the material cooling from outflows is likely to be a strong function of the host galaxy properties, hinting at the wide range of Lyman-$\alpha$ luminosities seen in rapidly star formation galaxies \citep{2015ApJ...803....6M}.

The last four panels of Figure \ref{dspectra} show the evolution of the density power spectra in the ULRIG cases.    Unlike the M82 runs, structure grows most rapidly in the ULIRG simulations between $3 \times 10^4$ and $3 \times 10^5$ years, because the high-mass fluxes in these models lead to rapid density-dependent cooling much earlier in the outflow evolution.  On the other hand, in the ULIRG simulations, this early cooling also causes the underdense regions to quickly drop to the $\approx 2 \times 10^4$K  point at which cooling and photo-heating are balanced (see Figure \ref{Fig:cool}).   Since this now happens are small radii, these fluctuations are not yet frozen into the medium.   Thus pressure difference begin to act to reverse the growth of structure, leading to density distributions at $3 \times 10^6$ years that are somewhat more homogeneous than in the M82 cases.

\section{Observability}

In Figure \ref{fig:coldr}, I show the radial distribution of $\left < n_{\rm cold}^2 \right>^{1/2},$ the rms number density of material between $10^4$K and $3 \times 10^4$K, a measure of the emission expected in low ionization state metal lines.   Note that although each simulation tracks a single comoving region 150 parsec on a side as it moves outward, this is a piece of an overall steady-state outflow that will continue for as long as the mass and energy input are sustained.   Thus at any given time, the entire radial distribution depicted in these figures would be expected to contribute to observed emission.  Moreover, as the models only include photoheating from the source galaxy and not the additional contribution from the metagalactic background or the flux from nearby galaxies, at large distances the density of gas with observable low-ionization state ions may be somewhat greater than shown here.   This means the value in Figure \ref{fig:coldr} should be thought of as a conservative lower limit on the  rms number density of observable cold material, especially at large distances.

The upper left panel of this figure shows the results from the fiducial M82 simulations with mass outflow rates of $10 M_\odot$ yr$^{-1},$ and various cooling processes included.  Here we see that  it is Compton cooling that has the largest impact on the radius at which gas cools efficiently.  This is due both to the suppression of atomic/ionic cooling and the enhancement of Compton cooling at short distances from the starburst.   On the other hand, the presence of dust has only a small impact on the cold gas properties, because it is quickly destroyed at the high densities found in this case.  

Moving to M82 models with the same energy flux ($\epsilon =1$) but lower mass fluxes, the impact  of atomic/ionic cooling and Compton cooling is dramatically reduced.  This is due to the fact that the outflows are launched  at higher temperatures and at higher velocities, such that these processes have more internal energy per particle to remove and less time to do so.  Thus these cases only cool below $3 \times 10^4$K at distances beyond $\approx 2$ kpc and at rms densities below $\approx 10^{-2}$cm$^{-3},$ meaning that their emission is likely to be very difficult to observe.   When the energy flux is reduced by a factor of two ($\epsilon =0.5$), cooling processes are substantial in the $\dot M_{\rm hot}  = 5M_\odot$ yr$^{-1}$ case, but minimal in the $\dot M_{\rm hot} =  2 M_\odot$ yr$^{-1}$ and $\dot M_{\rm hot} = 1 M_\odot$ yr$^{-1}$ cases, illustrating the strong sensitivity of the wind to initial conditions.

In the ULIRG case in which the star formation rate, mass outflow rate, and energy input rate are all boosted by a factor of ten, rapid cooling is a generic feature of all models.   Thus  in  the $\dot M_{\rm hot} = 50M_\odot$ yr$^{-1}$ cases  $\left < n_{\rm cold}^2 \right>^{1/2}$ approaches $\approx 10$ cm$^{-3}$ at distances just outside the driving radius, while in the $\dot M_{\rm hot} = 20M_\odot$ yr$^{-1}$ cases,  $\left < n_{\rm cold}^2 \right>^{1/2}$ approaches $\approx 1$ cm$^{-3}$ at distances of $\approx 500-800$ parsecs.  Note also that in the $\dot M_{\rm hot} = 50M_\odot$ yr$^{-1},$ $\epsilon = 0.5$ case the outflow shows a double peaked profile, due to the clumping of structure that occurs as $T$ drops rapidly through the unstable regime.
\begin{figure*}
\center{\includegraphics[width=170mm]{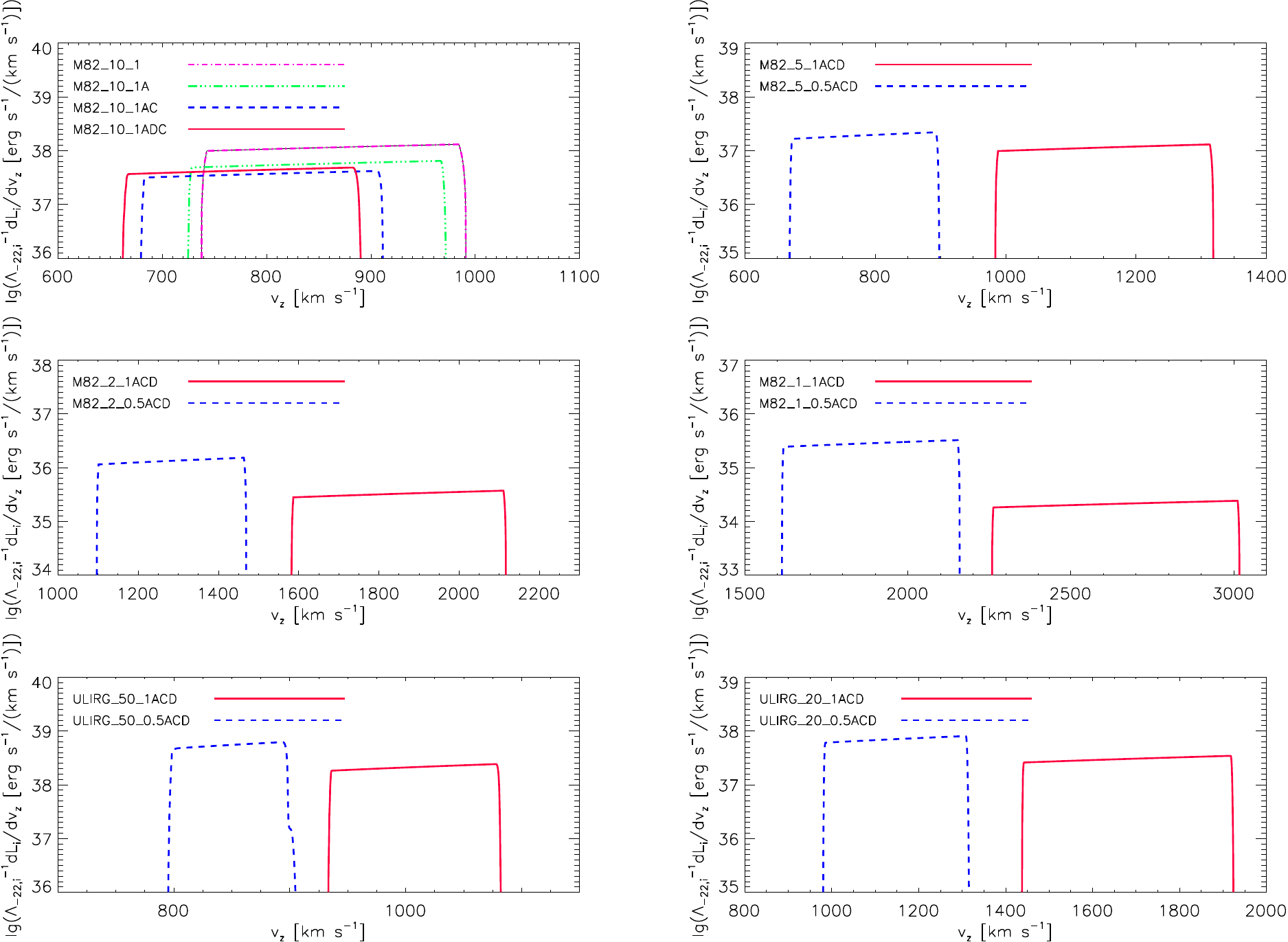}}
\caption{$\Lambda_{-22,i}^{-1} \frac{dL_i}{dv_z}$, the luminosity normalized by $\Lambda_{-22,i} \equiv \Lambda_{i} / (10^{-22} \, {\rm ergs} \, {\rm s}^{-1} {\rm cm}^3)$ as computed
 from a simple model of emission in which a biconical outflow covering $\Omega = \pi$ steradians is being observed directly along its axis (eq.\ \ref{eq:dLdvz}).  Models and lines are as in Figure \ref{fig:coldr}. \vspace{0.18in}}
\label{fig:coldv}
\end{figure*}

As a more direct measure of observability I also computed the approximate luminosity per line-of-sight sight velocity for a simple model in which observations are taken straight along the axis of the outflow, which encompasses $\pi$ steradians, $\pi/2$ on each side of the galaxy.  In this case, the luminosity in line $i$  per line-of-sight velocity is given by
\ba
\frac{dL_i}{dv_z} &=&  
\Lambda_{i}  2 \pi   \int_{300 \rm pc}^\infty dr r^2 \left < n_{\rm cold}^2 (r)\right> \nonumber\\
& & \qquad \qquad
\int^1_{3/4} d \cos \theta  \delta \left[\frac{v_z}{\cos \theta - v(r)} \right], \nonumber \\
&=& 
 \Lambda_{i} 2 \pi v_z
 \int_{v_z}^{4/3 v_z} \frac{dv}{v^2}  r^2(v) \frac{dr}{dv}(v) \left < n_{\rm cold}^2 (v)\right>, 
\label{eq:dLdvz}
\ea
where $\Lambda_{i},$ which has units of erg s$^{-1}$ cm$^3$ is the average power per unit volume emitted in a line of interest by a medium with an average cold gas density squared of  $\left < n_{\rm cold}^2\right>.$  Here a positive $v_z$ is defined as a velocity towards the observer, corresponding to a blueshift of the emission line, and typical values of $\Lambda_{i},$ are on the order of 10$^{-23}$ to 10$^{-22}$ ergs s$^{1}$ cm$^3$ \citep{Boehringer1989,Sutherland1993,Wiersma2009,Gnedin12}.

In Figure \ref{fig:coldv}, I plot $\Lambda_{-22,i}^{-1} \frac{dL_i}{dv_z}$ from each of the simulations, the luminosity normalized by $\Lambda_{-22,i} \equiv \Lambda_{i} / (10^{-22} \, {\rm ergs} \, {\rm s}^{-1} {\rm cm}^3).$  As in Figure \ref{fig:coldr}, the normalized luminosities shown here represent lower limits because none of these models includes a metagalatic background. The upper left panel of this figure shows the results from the fiducial M82 simulations with mass outflow rates of $10 M_\odot$ yr$^{-1}$ and various cooling processes included.  Here we see that while the inclusion of additional cooling processes has a large impact on the radial distance at which gas cools, it has a much smaller impact on the luminosity as a function of velocity.   This is due to the fact that although the additional processes cause the gas to cool at higher densities, they also cause the volume of the outflow in the $10^4$ to $3 \times 10^4$K range to decrease significantly.   Note also that the range of line-of-sight velocities over which outflowing gas is observed has very little to do with its radial distance, and the profiles are instead dominated by geometric effects. 

In M82 models with $\epsilon =1$  and lower mass fluxes, the relative inefficiency of atomic/ionic cooling and Compton cooling cause the cold gas to only be found at low densities and radial velocities approaching the maximum possible value of $2 v_\star.$  These effects cause the overall normalized luminosity to decrease and the line-of-sight velocity to increase rapidly as a function of $M_\odot$.   Thus the $5 M_\odot$ yr$^{-1}$ case is roughly 3 times fainter than the $10 M_\odot$ yr$^{-1}$ case, with a velocity offset that is 1.5 time as large, while the $2 M_\odot$ yr$^{-1}$ and $1 M_\odot$ yr$^{-1}$ cases are 100 and 1000 times fainter than the fiducial case, respectively, with velocity offsets that are 2 and 3 times as large.   Furthermore, while decreasing  the energy input ($\epsilon =0.5$) boosts the luminosities and decreases the distances somewhat, the very strong trend with mass flux means that the low $M_\odot$ outflows are likely to remain undetectable.  Note that the strong dependence of the total luminosity of the outflow to the properties of the host is consistent with the large dynamic range of Ly$\alpha$ luminosities observed in outflowing galaxies \citep{2015ApJ...803....6M}.

It also means that moving to higher mass fluxes, as in the ULIRG cases, yields overall luminosities that are much higher than in the M82 cases.  Furthermore, the strong cooling in these models, driven both by  the higher densities as well as the higher background fluxes, means that the cold material is found at velocities well below $2 v_\star,$ the maximum velocity without cooling included.  Thus the $\approx 1000$ km/s maximum velocities found in the ($\epsilon =0.5$) cases are consistent with maximum Ly$\alpha$ and [OIII] $\lambda$5007 blueshifts observed in ULIRGs \citep{2015ApJ...803....6M}  although the velocities are slightly higher than typically found, while the luminosities are slightly lower, suggesting that models with $\epsilon < 0.5$ may provide even better matches to the observations.

On the other hand, the velocities in the models are significantly   greater than observed in absorption line measurements of Na I \citep[e.g.][]{2005ApJ...621..227M,2006ApJ...647..222M,2005ApJS..160..115R} and OH \citep[e.g.][]{2013ApJ...776...27V}.    One possibility is that the molecular and NaI absorption arise from cold clouds that are accelerated and shredded at small distances by the hot wind.  This entrainment and heating of this material would then lead to additional radiative losses within the driving regime, such that smaller $\epsilon$ values would be favored, leading to strong Ly$\alpha$ and [OIII] emission as the hot wind itself cooled as it expanded into the circumgalactic medium.  Further modeling  and comparisons with ULIRG observations are needed to properly assess this scenario.
 
\section{Conclusions}

Starburst-driven galaxy outflows play a key role in the history of structure formation and are well observed at a variety of redshifts.   However, our ability to interpret these observations is strongly hampered by the fact that the $\approx 10^4$K material that is easiest to observe is also the most difficult to model theoretically. While it is often assumed that this material arises from cold clouds accelerated by the hot wind,  detailed models of cold-cloud hot-wind interactions show that instabilities and evaporation are likely to shred clouds well before they are significantly accelerated.   A second possibility is that the cold material is formed directly from the cooling of hot wind material, cooling it into the observable range after it is already moving at high radial velocities.  

To test this scenario, I have carried out a suite of  fully-three dimensional simulations of initially-hot material ejected by starburst-driven galaxy outflows.
The simulations are conducted in a comoving frame that moves with the material, and,  building on the cooling routines presented in \cite{2016arXiv160902561F}, they simulations include atomic/ionic cooling that accounts for the ionizing background, as well as Compton cooling, and cooling from dust grains.   I find that  atomic/ionic cooling is significantly reduced due to photoionization, particularly at small radii around galaxies with higher star formation rates.   Instead,  it is Compton scattering with photons from the host galaxy that has the largest impact, matching or exceeding temperature losses due to all other mechanisms including adiabatic expansion. Compton cooling becomes particularly important when the SFR is large, such that the galaxy is luminous, and when the mass flux is large, such that the outflowing material spends a long time near the galaxy.   

The most important role of atomic/ionic cooling, on the other hand, is enhancing inhomogeneities in the medium. While it is subdominant at high temperatures, these processes overtake the adiabatic and Compton cooling rates once the overall temperature has dropped to below $\approx 10^6$K.  Because collisional cooling occurs most rapidly in the densest regions, the pressure within them drops significantly with respect to their surroundings, leading to further increases in density. This rapidly amplifies densities inhomogeneities by roughly an order of magnitude until the medium as a whole reaches $\approx 10^4$K and the sharp decrease in sound speed freezes out further growth of structure.   

On the other hand, while dust is an extremely efficient coolant, it has little effect on outflow evolution, as  it is rapidly destroyed in the extreme conditions found in most galaxy outflows.   In fact the only case that dust is preserved throughout the full outflow evolution is the fastest moving, lowest density case I considered. This suggests that small starbursts with modest mass outflow rates, rather than large dusty starbursts,  may be the most efficient  at polluting the circumgalactic medium with dust.

The simulation results can be used to construct a simple steady-state model of the observed UV/optical emission from each outflow, adopting a simple approximation that the outflow is observed along its central axis.  In this case, the radial distance at which the outflow cools has only a minor impact on the observations, as models that cool more quickly have higher cold gas densities but also much smaller cold-gas volumes.   In fact, the spread in the of line-of-sight velocities of the cold material has very little to do with its radial distance, and the profiles are instead dominated by geometric effects.  

On the other hand, the total luminosities and maximum velocities are strong functions of the properties of the host, which is consistent with the large dynamic range of Ly$\alpha$ and [OIII] luminosities observed in outflowing ULIRGs.   In fact, the luminosities and maximum velocities found in the models with significant energy losses ($\epsilon = 0.5$)  are consistent with such observations, although the velocities are greater than observed in NaI and OH absorption line studies.   It may be that absorption lines probe entrained cold material, while emission lines probe clumps of cold material condensing out the hot wind, a possibility that deserves further theoretical study.

\section*{Acknowledgments} 

I would like to thank Andrea Ferrara, Crystal Martin, and Sylvain Veilleux for insightful discussions that greatly improved this paper. This work was supported by the National Science Foundation under grant AST14-07835 and PHY11-25915 and NASA theory grant NNX15AK82G. I thank the Simons Foundation and the organizers of the workshop {\em Galactic Winds: Beyond Phenomenology} (J. Kollmeier and A. Benson) and the Kavli Institute for Theoretical Physics and the organizers of the program {\em The Cold Universe} (P. Caselli, A. Ferrara, M. Ouchi, R. Schneider, and J. Tan), for discussions that helped to inspire and refine the work carried out here.  I would also like to thank the Texas Advanced Computing Center (TACC) at The University of Texas at Austin, and the Extreme Science and Engineering Discovery Environment (XSEDE) for providing HPC resources via grant TG-AST130021 that have contributed to the results reported within this paper.

\vspace{1.2in}


\bibliographystyle{../style/apj}


\end{document}